\documentclass[%
reprint,
superscriptaddress,
amsmath,amssymb,
aps,
prb,
]{revtex4-2}

\usepackage{graphicx}% Include figure files
\usepackage{dcolumn}% Align table columns on decimal point
\usepackage{bm}% bold math
\usepackage{hyperref}% add hypertext capabilities
\hypersetup{hypertex=true, colorlinks=true, linkcolor=blue, anchorcolor=blue, citecolor=blue, urlcolor=blue}
\usepackage[mathlines]{lineno}% Enable numbering of text and display math
\begin{document}
	
\preprint{APS/123-QED}
	
\title{Topology reconstruction for asymmetric systems by isomorphic mapping or perturbation approximation}% Force line breaks with \\
%\thanks{jiangxunya@fudan.edu.cn}%
	
\author{Yunlin Li}
\affiliation{%
	Department of Illuminating Engineering and Light Sources, School of Information Science and Engineering, Fudan University, Shanghai 200433, China}%
\affiliation{%
	Engineering Research Center of Advanced Lighting Technology, Fudan University, Ministry of Education, Shanghai 200433, China}%

\author{Jingguang Chen}
\affiliation{%
	State Key Laboratory of Surface Physics, Key Laboratory of Micro- and Nano-Photonic Structures (Ministry of Education) and Department of Physics, Fudan University, 200433 Shanghai, China}%

\author{Xingchao Qi}
\affiliation{%
	Institute of Future Lighting, Academy for Engineering and Technology, Fudan University, Shanghai 200433, China}%
\affiliation{%
	Engineering Research Center of Advanced Lighting Technology, Fudan University, Ministry of Education, Shanghai 200433, China}%

\author{Langlang Xiong}
\affiliation{%
	Institute of Future Lighting, Academy for Engineering and Technology, Fudan University, Shanghai 200433, China}%
\affiliation{%
	Engineering Research Center of Advanced Lighting Technology, Fudan University, Ministry of Education, Shanghai 200433, China}%
	
\author{Xianjun Wang}
\affiliation{%
	Department of Illuminating Engineering and Light Sources, School of Information Science and Engineering, Fudan University, Shanghai 200433, China}%
\affiliation{%
	Engineering Research Center of Advanced Lighting Technology, Fudan University, Ministry of Education, Shanghai 200433, China}%

\author{Yufu Liu}
\affiliation{%
	Department of Illuminating Engineering and Light Sources, School of Information Science and Engineering, Fudan University, Shanghai 200433, China}%
\affiliation{%
	Engineering Research Center of Advanced Lighting Technology, Fudan University, Ministry of Education, Shanghai 200433, China}%
	
\author{Fang Guan}
\affiliation{%
	State Key Laboratory of Surface Physics, Key Laboratory of Micro- and Nano-Photonic Structures (Ministry of Education) and Department of Physics, Fudan University, 200433 Shanghai, China}%
	
\author{Lei Shi}
\affiliation{%
	State Key Laboratory of Surface Physics, Key Laboratory of Micro- and Nano-Photonic Structures (Ministry of Education) and Department of Physics, Fudan University, 200433 Shanghai, China}%

\author{Xunya Jiang}%
\email{jiangxunya@fudan.edu.cn}
\affiliation{%
	Department of Illuminating Engineering and Light Sources, School of Information Science and Engineering, Fudan University, Shanghai 200433, China}%
\affiliation{%
	Engineering Research Center of Advanced Lighting Technology, Fudan University, Ministry of Education, Shanghai 200433, China}%
\affiliation{%
		Institute of Future Lighting, Academy for Engineering and Technology, Fudan University, Shanghai 200433, China}%
	
\begin{abstract}
The systems without symmetries, e.g. the spatial and chiral symmetries, are generally thought to be improper for topological study and no conventional integral topological invariant can be well defined. In this work, with multi-band asymmetric Rice-Mele-like systems as examples, for the first time we show that the topology of all gaps can be reconstructed by two general methods and topological origin of many phenomena are revealed. A new integral topological invariant, i.e. the renormalized real-space winding number, can properly characterize the topology and bulk-edge correspondence of such systems. For the first method, an isomorphic mapping relationship between a Rice-Mele-like system and its chiral counterpart is set up, which accounts for the topology reconstruction in the half-filling gaps. For the second method, the Hilbert space of asymmetric systems could be reduced into degenerate subspaces by perturbation approximation, so that the topology in subspaces accounts for the topology reconstruction in the fractional-filling gaps. Surprisingly, the topology reconstructed by perturbation approximation exhibits extraordinary robustness since the topological edge states even exist far beyond the weak perturbation limit. We also show that both methods can be widely used for other asymmetric systems, e.g. the two-dimensional (2D) Rice-Mele systems and the superconductor systems. At last, for the asymmetric photonic systems, we predict different topological edge states by our topology-reconstruction theory and experimentally observe them in the laboratory, which agrees with each other very well. Our findings open a door for investigating new topological phenomena in asymmetric systems by various topological reconstruction methods which should greatly expand the category of topology study.
\end{abstract}
	
\maketitle
	
%\tableofcontents
	
\section{\label{sec:level1}INTRODUCTION}
Topological band theory has attracted much attention in both quantum and classic systems \cite{RevModPhys.88.021004,RevModPhys.83.1057,RevModPhys.91.015006,kim2020recent,lu2014topological,wang2017topological}. One of the most fascinating properties of topological phases is the robustness of topological edge states which shed light on the development of novel highly integrated devices \cite{PhysRevLett.100.013905,Ota:19}. The topological invariant defined in the bulk can predict the topological phases and the number of topological edge states through the bulk-edge correspondence. Since the discovery of tenfold classification in topological insulators and superconductors \cite{PhysRevB.78.195125,RevModPhys.88.035005} and the topological crystalline insulators \cite{PhysRevLett.106.106802}, it is widely believed that the nontrivial topology and the unique characteristics of topological edge states are related to certain symmetries. For example, in 1D systems, the topology is generally protected by chiral symmetry \cite{PhysRevB.98.024205,RevModPhys.91.015005,PhysRevB.90.125143,LI2021104837} which can be described by winding number \cite{PhysRevB.96.125418,PhysRevB.90.014505,PhysRevB.98.024205,PhysRevResearch.3.013148,LI2021104837}, or by spatial-inversion symmetry \cite{PhysRevB.96.245115,AshortCourse} which can be described by quantized Zak phase \cite{PhysRevB.96.245115,PhysRevLett.62.2747,RevModPhys.82.1959,PhysRevX.4.021017}. Similarly, for higher-order topology in 2D systems, the topological quadrupole insulators are protected by spatial symmetry \cite{PhysRevB.96.245115,science.aah6442} while the novel $\mathbb{Z}$-classified higher-order topological insulators require chiral symmetry \cite{PhysRevLett.128.127601,PhysRevLett.131.157201}.

A fundamental drawback of the symmetry-based topological theory is obvious and very deep, e.g. the periodic systems without certain symmetries are generally not considered in the category for topology study. That is to say, for systems without certain symmetries, it is rather tricky to analyze the topological origin of some phenomena, not to mention a well-defined topological invariant. On the other hand, recent studies in asymmetric systems \cite{PhysRevB.107.115401,PhysRevB.106.085109,PhysRevA.99.013833,WOS:000974238700002,yan2023topological} do exhibit clear signs of band topology as well as some novel phenomena such as non-zero edge states and topological phase transitions without gap closing and reopening. Obviously, the traditional symmetry-based topological theories are not suitable for these cases. However, to the best of our knowledge, little effort has been made to extend the symmetry-based topological theory into periodic systems without symmetries, especially, the general methods for such extending. Even more, the breakthrough of such extending can not only greatly expand the category of topological studies, but also reveal that some basic concepts that are generally thought to be irrelevant to the topological studies, e.g. the perturbation approximation, can take important roles in topological studies.

\begin{figure}[t]
	\centering
	\includegraphics[width=1\linewidth]{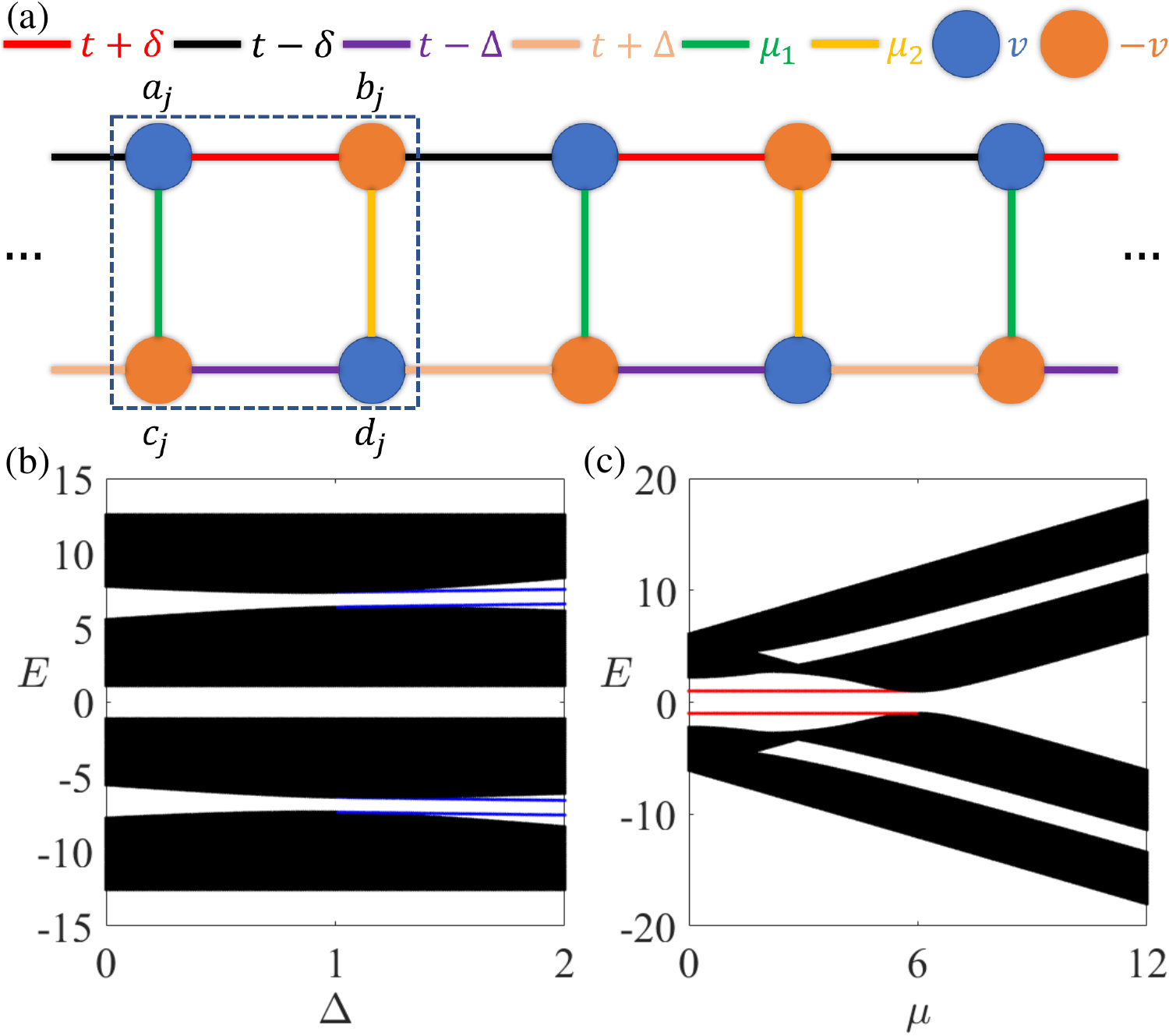}
	\caption{(a) Schematic of Rice-Mele ladder. The unit cell is marked by dashed box. (b) Band structure versus $\Delta$ with $t=3$, $\delta=1$, $\mu_1=6$, $\mu_2=7$, $v=1$. (c) Band structure versus $\mu_1=\mu_2=\mu$ with $t=3$, $\delta=2$, $\Delta=1$, $v=1$. Both (b) and (c) are calculated by the system with $50$ unit cells with OBC.} \label{fig:1}
\end{figure}

In this work, we introduce a novel perspective for the topology study of asymmetric systems, in which two general methods can be introduced to reconstruct the topology for such systems. The first method is by a strict mathematical transformation, which is an isomorphic mapping for the wave functions and eigenvalues from the Rice-Mele-like system to its chiral counterpart. Based on isomorphic mapping, we show that the nontrivial topology for the half-filling gap and the bulk-edge correspondence can be precisely described by a new topological invariant, i.e. the renormalized real-space winding number, which can be well defined. The second method is by a physical transformation based on perturbation approximation. This is very surprising since the perturbation approximation is generally thought to be irrelevant to the topology. We show that a new topology structure from the degenerate subspace can emerge after perturbation approximation and some interesting phenomena, e.g. the edge states in the fractional-filling gaps of asymmetric systems, are dominated by the nontrivial topology from the degenerate subspace. Even more, such approximation-based topology and edge states exhibit extraordinary robustness since they can go far beyond the weak perturbation limit. The perturbation-based topology can also be characterized by the renormalized real-space winding number with redefined sublattice. The Rice-Mele-like systems are chosen as 1D multi-band asymmetric examples to show the effectiveness of our methods and new related topological phenomena are investigated. Besides the Rice-Mele-like systems, we show that our methods can also be utilized in other systems, such as 2D Rice-Mele model and 1D dimerized superconducting model. At last, we introduce these methods to the classical systems, e.g. the asymmetric photonic crystals(PhCs) whose effective Hamiltonian is similar as that of Rice-Mele ladder. The topological phases and edge states of PhCs are also identified by our methods based on the photonic tight-binding model theory. Experiments are done for the asymmetric PhCs and the topological edge states at the boundary between PhCs with different topological phases are observed in the laboratory, which agrees very well with our theoretical prediction. This work opens a door for the study of topology in asymmetric systems and a great amount of new topological phenomena can be expected. We demonstrate that the topological characteristics of asymmetric systems can still be described rather precisely as long as the topology can be reconstructed, which could be a new paradigm for topology study.

\section{\label{sec:level1}MODEL AND TOPOLOGICAL INVARIANTS}

\subsection{\label{sec:level2}Model}

In this work, we would like to take the Rice-Mele ladder as the first example to illustrate our topology reconstruction methods in asymmetric systems. Shown in Fig. \ref{fig:1}(a), our Rice-Mele ladder is a natural extension from the 1D Rice-Mele model \cite{PhysRevLett.49.1455} to a ladder structure. The model could also be realized by introducing staggered non-zero onsite energies to SSH ladder \cite{PhysRevB.98.024205}. The Hamiltonian could be written as:
\begin{eqnarray} \label{eq1}
	\begin{aligned}
		\mathcal{H}&=\sum_{j}[(t+\delta)b_j^{\dagger}a_j+(t-\delta)a_{j+1}^{\dagger}b_j+(t-\Delta)d_j^{\dagger}c_j
		\\
		&+(t+\Delta)c_{j+1}^{\dagger}d_j+\mu_1 a_j^{\dagger}c_j+\mu_2 b_j^{\dagger}d_j+H.c.]
		\\
		&+va_j^{\dagger}a_j-vb_j^{\dagger}b_j-vc_j^{\dagger}c_j+vd_j^{\dagger}d_j.
	\end{aligned}
\end{eqnarray}
where $\alpha$ and $\alpha^{\dagger}$ are the annihilation and creation operators at site $\alpha$, $\alpha=a,b,c,d$. Both chiral and spatial-inversion symmetries are broken for the Rice-Mele ladder. When we set the onsite energy $v=0$, the model degenerates to the SSH ladder whose topology at half-filling is protected by chiral symmetry \cite{PhysRevB.96.125418,PhysRevB.98.024205,WOS:000497712600004,WOS:000570035500003,ryu2022topological}.

The Bloch Hamiltonian is given by:
\begin{eqnarray} \label{eq2}
	\mathcal{H}(k)=
	\begin{bmatrix}
		vI_2 & M(k)\\
		M(k)^{\dagger} & -vI_2
	\end{bmatrix},
\end{eqnarray}
here $M(k)=\begin{bmatrix}
	\mu_1 & (t+\delta)+(t-\delta)e^{ik}\\
	(t-\Delta)+(t+\Delta)e^{-ik} & \mu_2
\end{bmatrix}$ and $I_n$ stands for the $n \times n$ identity matrix. The Hamiltonian of Eq. \eqref{eq2} can be thought of as the generalization of the Hamiltonian of common Rice-Mele model \cite{PhysRevLett.49.1455} since both $M$ and $vI_2$ are matrices.

The eigenstates of the Rice-Mele ladder versus the parameters $\Delta$ and $\mu$ with the open boundary condition (OBC) are exhibited in Fig. \ref{fig:1}(b) and (c), respectively. Besides the band-gap structure, when the parameters pass certain critical values the in-gap edge states appear at the narrowest points of the gap, which we will show that the phenomena are typical signs of topological phase transition for systems without chiral and spatial-inversion symmetries. As a four-band system, the Rice-Mele ladder possesses three gaps and all of them are candidates for our topology study. To avoid any possible misunderstanding, the edge states in the $2$nd gap, corresponding to the half-filling gap, are marked by red lines, while the edge states in the $1$st and $3$rd gap, corresponding to the other fractional-filling gaps, are marked by blue lines. We will see that the topological origins of these phenomena signified by different colors are different and they need different topology reconstruction methods.

\subsection{\label{sec:level2}Topological Invariants}
In the study of topological band theory, the importance of topological invariant is self-evident. As mentioned above, the definition of most conventional topological invariants for 1D systems, e.g. the Zak phase and the $k$-space winding number, relies heavily on system symmetries which makes them more difficult for our asymmetric systems. Also, the quantization of these conventional topological invariants is another severe problem for asymmetric systems. Very recently, real-space winding number \cite{PhysRevB.103.224208,PhysRevLett.128.127601} shows its power in the topology study. Although it is initially developed for describing topological properties of random systems without translational symmetry \cite{PhysRevB.103.224208,WOS:000839909500001,PhysRevB.106.134201}, it also offers a potential method to study the topology of period systems without spatial-inversion and chiral symmetries.

The real-space winding number with periodic boundary condition(PBC) is given by \cite{PhysRevB.103.224208}:
\begin{eqnarray} \label{eq3}
	w = \frac{1}{2\pi i} \mathrm{Tr} \left [ \mathrm{log}(\chi_{A}\chi_{B}^{-1} ) \right ],
\end{eqnarray}
where
\begin{eqnarray} \label{eq4}
	\chi_{\sigma} = U_{\sigma}^{-1}\Gamma_{\sigma}\chi\Gamma_{\sigma}U_{\sigma},
\end{eqnarray}
for $\sigma=A,B $. The $i$th column in $U_{\sigma}$ is the $\sigma$ sector of the $i$th eigenstate below the gap under consideration. $\Gamma_{\sigma} = {\textstyle \sum_{l,\alpha \in \sigma} | l,a \rangle \langle l,a |} $ is the projector operator of sublattice $\sigma$. The position operator $\chi$ reads:
\begin{eqnarray} \label{eq5}
	\chi=\sum_{l}e^{i\frac{2\pi}{N}l} \left (| l,A \rangle \langle l,A |+| l,B \rangle \langle l,B |\right).
\end{eqnarray}
Unless otherwise specified, in this work we calculate the real-space winding number $w$ for the $50$-unit-cell system with PBC. We  prove that the necessary condition for a system to be described by quantized real-space winding number is [Appendix \ref{A}]:
\begin{eqnarray} \label{eq6}
U_AU_A^{\dagger}=U_A^{\dagger}U_A=U_BU_B^{\dagger}=U_B^{\dagger}U_B=0.5.
\end{eqnarray}
Just like its $k$-space counterpart which is defined by the structure of Bloch Hamiltonian, the real-space winding number also requires chiral symmetry [Appendix \ref{A}] and can not be directly used for our theory. However, as the real-space winding number is calculated directly based on eigenstates, it shows more flexibility considering the fact that the eigenstates can be redefined or renormalized, etc. This property enables us to define new topological invariant with some improvements which will be presented in the following sections.
		
\section{\label{sec:level1}Topology Reconstruction via Isomorphic Mapping}

In this section, we would like to introduce a general method of topology reconstruction for the asymmetric systems whose Hamiltonian has a similar form of Eq. \eqref{eq2}, which is referred to as the Rice-Mele-like Hamiltonian in this paper. The method is based on isomorphic mapping, by which a one-to-one mapping between the wave functions of Rice-Mele-like Hamiltonian and those of a new Hamiltonian with chiral symmetry can be found, so that the new topology of Rice-Mele-like Hamiltonian can be reconstructed for \emph{the gap around zero-energy}, corresponding to the half-filling gaps generally. By this method, we can also define a new topological invariant for such systems, namely the renormalized real-space winding number, which can clearly show the topological phase transition at critical point by quantized jumping and the topological edge states appearing. The method can be easily extended to higher dimensional systems, such as the 2D Rice-Mele systems.

\subsection{\label{sec:level2}The General Method of Topology Reconstruction by Isomorphic Mapping}
Similar as Eq. \eqref{eq2}, the real-space Hamiltonian of Rice-Mele-like models takes the general form as:
\begin{eqnarray} \label{eq7}
	\mathcal{H}_{RM}=\begin{bmatrix}
		vI_N & h \\
		h^{\dagger} & -vI_N
	\end{bmatrix}.
\end{eqnarray}
Its eigen-equation
\begin{eqnarray} \label{eq8}
	\mathcal{H}_{RM}\begin{bmatrix}
		\varphi_{An} \\
		\varphi_{Bn}
	\end{bmatrix}=E_n\begin{bmatrix}
		\varphi_{An} \\
		\varphi_{Bn}
	\end{bmatrix}
\end{eqnarray}
can be decomposed into two equations:
\begin{eqnarray}
	v\varphi_{An}+h\varphi_{Bn}=E _n\varphi_{An} , \label{eq9} \\
	h^{\dagger}\varphi_{An}-v\varphi_{Bn}=E _n\varphi_{Bn} \label{eq10} .
\end{eqnarray}
Pre-multiply Eq. \eqref{eq9} by $h^{\dagger}$ and Eq. \eqref{eq10} by $h$, after some simple calculation, we can obtain two decoupled equations for $\varphi_{An}$ and $\varphi_{Bn}$:
\begin{eqnarray} \label{eq11}
	\left\{\begin{matrix}
		hh^{\dagger}\varphi_{An}=(E_n^2-v^2)\varphi_{An}\\
		h^{\dagger}h\varphi_{Bn}=(E_n^2-v^2)\varphi_{Bn}
	\end{matrix}\right.
\end{eqnarray}
Eq. \eqref{eq11} ensures the orthogonality of eigenstates $\varphi_{\sigma n}$, $\sigma=A,B$. However, since chiral symmetry is broken in Rice-Mele-like system, the normalization condition $\langle \psi_{An} | \psi_{Bn} \rangle = \langle \psi_{Bn} | \psi_{Bn} \rangle = 0.5$ is broken. That is to say, Eq. \eqref{eq6} isn't satisfied for $U_\sigma$ and the real-space winding number is unavailable.

On the other hand, the following eigen-equation can also obtain the form of Eq. \eqref{eq11}:
\begin{eqnarray} \label{eq12}
	\mathcal{H}_{Chiral}
	\begin{bmatrix}
		\bar{\varphi}_{An} \\
		\bar{\varphi}_{Bn}
	\end{bmatrix}=\sqrt{E_n^2-v^2}
	\begin{bmatrix}
		\bar{\varphi}_{An} \\
		\bar{\varphi}_{Bn}
	\end{bmatrix},
\end{eqnarray}
where
\begin{eqnarray} \label{eq13}
	\mathcal{H}_{Chiral}=
	\begin{bmatrix}
		0 & h \\
		h^{\dagger} & 0
	\end{bmatrix}
\end{eqnarray}
exhibits the characteristics of chiral symmetry, i.e. the chiral operator $\Gamma$ anti-commutes with $\mathcal{H}_{Chiral}$:
\begin{eqnarray} \label{eq14}
	\Gamma\mathcal{H}_{Chiral}\Gamma^{\dagger}=-\mathcal{H}_{Chiral} \ \text{with} \ \Gamma=\begin{bmatrix}
		I_N&0 \\
		0&-I_N
	\end{bmatrix}.
\end{eqnarray}
That is to say, $\bar{\varphi}_{\sigma n}$ is normalized to $0.5$. Physically, the system of Eq. \eqref{eq8} with Hamiltonian $\mathcal{H}_{RM}$ and the system of Eq. \eqref{eq12} with Hamiltonian $\mathcal{H}_{Chiral}$ are quite different. However, from the view of wave functions and eigenvalues, one can set up a strict mapping rule:
\begin{eqnarray} \label{eq15}
	\left\{\begin{matrix}
		\begin{bmatrix}
			\varphi_{An} \\
			\varphi_{Bn}
		\end{bmatrix}\longrightarrow
		\begin{bmatrix}
			\bar{\varphi}_{An} \\
			\bar{\varphi}_{Bn}
		\end{bmatrix}
		\\\\
		E_n\longrightarrow\sqrt{E_n^2-v^2}
	\end{matrix}\right.
\end{eqnarray}
Also, comparing Eq. \eqref{eq11} and Eq. \eqref{eq12}, we can see that the difference between $\bar{\varphi}_{\sigma n}$ and $\varphi_{\sigma n}$ is up to a normalization factor:
\begin{eqnarray} \label{eq16}
	\begin{bmatrix}
			\bar{\varphi}_{An} \\
			\bar{\varphi}_{Bn}
		\end{bmatrix}=
		\begin{bmatrix}
			C_{An}\varphi_{An} \\
			C_{Bn}\varphi_{Bn}
		\end{bmatrix},
\end{eqnarray}
where $C_{\sigma n}$ can be calculated simply based on the normalization condition of $\bar{\varphi}_{\sigma n}$:
\begin{eqnarray} \label{eq17}
	C_{\sigma n} = \frac{1}{\sqrt{2}\langle \varphi_{\sigma n} | \varphi_{\sigma n} \rangle}.
\end{eqnarray}
At here, we emphasize two points. The first point is that $C_{An}$ and $C_{Bn}$ are obtained independently and they could be quite different. This is quite anti-intuitive if we keep in mind that the original normalization condition is $ \langle \varphi_{An} | \varphi_{An} \rangle + \langle \varphi_{Bn} | \varphi_{Bn} \rangle=1$. The second point is that the derivation above is valid for both PBC and OBC since the difference between PBC and OBC is just on the detail of $h$. Apparently, the normalization coefficients $C_{\sigma n}$ wouldn't change the exponential property of edge states with OBC. So, we conclude that the system with $\mathcal{H}_{RM}$ and the system with $\mathcal{H}_{Chiral}$ have the one-to-one-mapping edge states and the properties of edge states are quite similar except for a factor difference for every sublattice.

Next, we would like to consider the corresponding topological invariant. As the real-space winding number is well-defined for the renormalized eigenstates $\bar{\varphi}_{\sigma n}$, we can construct a renormalized $\bar{U}_\sigma$ matrix via $\bar{\varphi}_{\sigma n}$:
\begin{eqnarray} \label{eq18}
	\bar{U}_\sigma(i,j) = \bar{\varphi}_{\sigma j}^{(i)} = \frac{1}{\sqrt{2}\langle \varphi_{\sigma j} | \varphi_{\sigma j} \rangle}\varphi_{\sigma j}^{(i)},
\end{eqnarray}
where $j$ is the index of state, $i$ is the index of site. One can expect that Eq. \eqref{eq6} is satisfied again for $\bar{U}_\sigma$ which can be thought of as the matrices from ${U}_\sigma$ after renormalization, so that new topological invariant can be calculated based on it. Here, we refer to the new topological invariant calculated by $\bar{U}_\sigma$ matrices, instead of ${U}_\sigma$, as \emph{the renormalized real-space winding number}. From the new topological invariant, we can further conclude that the Rice-Mele-like system shares the same topology as its chiral counterpart.

\begin{figure}[t]
	\centering
	\includegraphics[width=1\linewidth]{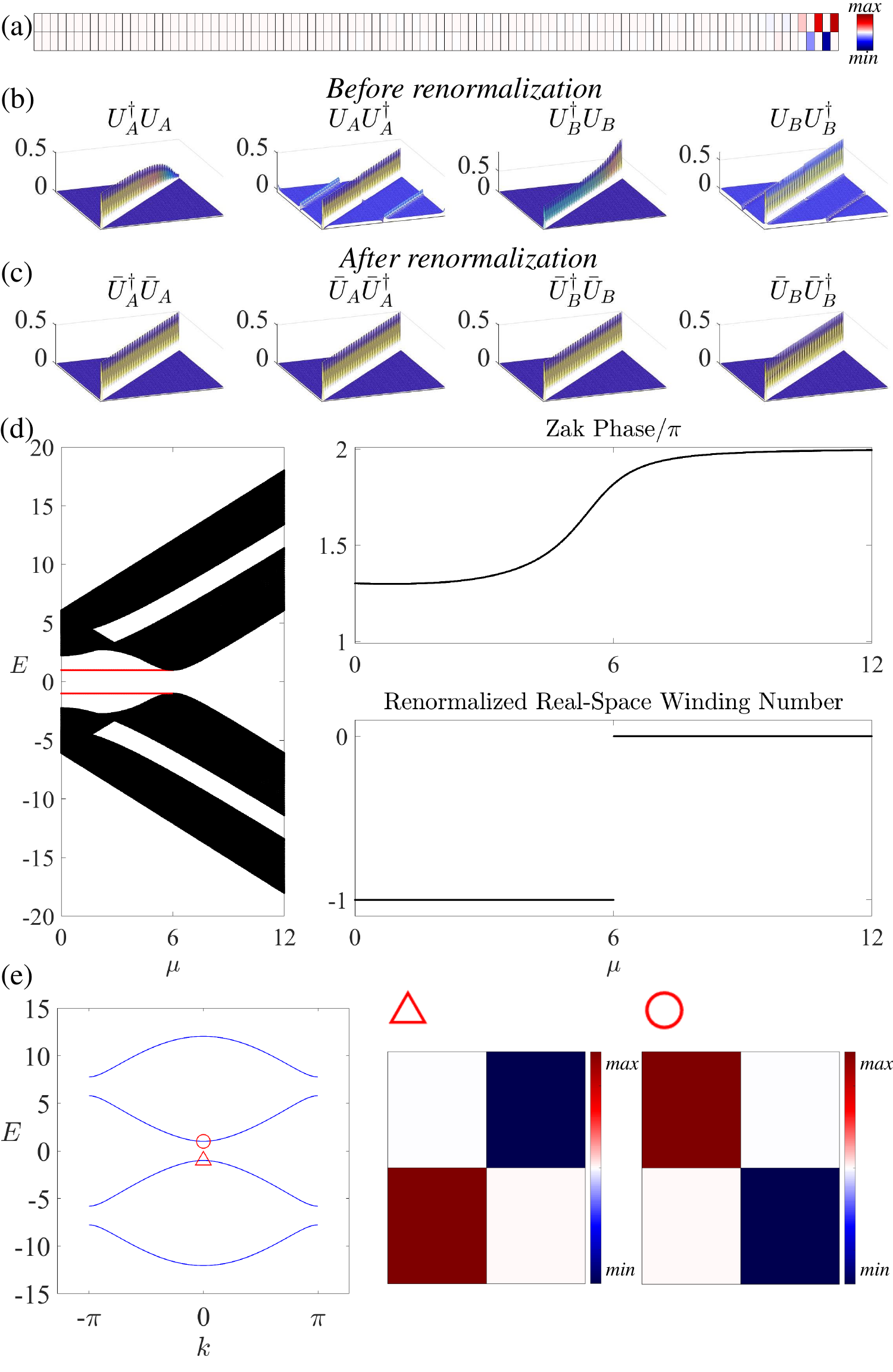}
	\caption{(a) The spatial distribution of edge state in the half-filling gap with $t=3$, $\delta=1$, $\Delta=2$, $\mu_1=5$, $\mu_2=4$, $v=1$. (b-c) The distributions of the four product matrices $U_AU_A^{\dagger}$, $U_A^{\dagger}U_A$, $U_BU_B^{\dagger}$, $U_B^{\dagger}U_B$ calculated with or without renormalization, respectively. (d) The renormalized real-space winding number precisely predicts the topology of Rice-Mele ladder in the half-filling gap while the conventional topological invariant Zak phase fails. (e) The spatial distribution of the $2$nd and $3$rd Bloch eigenstates with $k=0$ at the topological phase transition point. Both of them are chiral eigenstates. } \label{fig:2}
\end{figure}

Obviously, our Rice-Mele ladder with Hamiltonian of Eq. \eqref{eq2} serves as a typical example of Rice-Mele-like systems with $N=2$. For the four intra-cell sites marked by $a,b,c,d$ in Fig. \ref{fig:1}(a), we define site $a,d$ as sublattice $A$ and $b,c$ as sublattice $B$. As we expect, a pair of topological edge states can be found in the second gap around zero energy. The spatial distribution of the edge state shown in Fig. \ref{fig:2}(a) exhibits the zig-zag feature, which can be regarded as typical chiral characteristics in a way that only one of the sublattices is occupied.

Meanwhile, for Rice-Mele ladder, the original $U_\sigma$ matrix calculated by $\varphi_{\sigma n}$ and the renormalized $\bar{U}_\sigma$ matrix calculated by $\bar{\varphi}_{\sigma n}$ are shown in Fig. \ref{fig:2}(b) and (c), respectively. One can see that while $U_A^{\dagger}U_A$ and $U_B^{\dagger}U_B$ are still diagonal matrices, their diagonal elements are no longer $0.5$. As a result, the off-diagonal elements are introduced into $U_AU_A^{\dagger}$, $U_BU_B^{\dagger}$. On the other hand, all of the four product matrices $\bar{U}_A\bar{U}_A^{\dagger}$, $\bar{U}_A^{\dagger}\bar{U}_A$, $\bar{U}_B\bar{U}_B^{\dagger}$, $\bar{U}_B^{\dagger}\bar{U}_B$ are diagonal to $0.5$, which satisfies Eq.(6). Therefore, we can calculate the renormalized real-space winding number, which is shown in Fig. \ref{fig:2}(d). Clearly, our new topological invariant can precisely predict the topological phase of the Rice-Mele ladder in the half-filling gap. By comparison, the conventional symmetry-based Zak phase fails to establish the bulk-edge correspondence.

Moreover, from Eq. \eqref{eq15} and Eq. \eqref{eq16}, two general properties can be found at the topological phase transition point of Rice-Mele-like systems. First, as shown in Fig. \ref{fig:2}(d), the gap around zero energy will not close at the topological phase transition point and achieve its minimum width, which is $2v$ according to Eq.\eqref{eq15}. Second, a pair of gap-edge eigenstates at the minimum gap width, which are signified by red triangle and circle in Fig. \ref{fig:2}(e), are \emph{the chiral eigenstates}, i.e. the eigenstates of the chiral operator $\Gamma$. The derivation of the second property can be seen in Appendix \ref{B}. Both properties can be widely used to identify the topological phase transition of Rice-Mele-like systems, especially for photonic systems which will be discussed later in Section \uppercase\expandafter{\romannumeral5}.

\subsection{\label{sec:level2}Topology Reconstruction in Higher-dimensional Rice-Mele-like Systems}

Since our method doesn't rely on any fine detail of matrix $h$ in Eq. \eqref{eq7}, it also shows potential in various systems beyond 1D. To demonstrate the power of our methods, in this section we would like to generalize the topology reconstruction into higher dimension.

In recent seminal works \cite{PhysRevLett.128.127601,PhysRevLett.131.157201}, it is shown that typical 2D chiral systems, such as the well-known Benalcazar-Bernevig-Hughes (BBH) model \cite{PhysRevB.96.245115,science.aah6442}, support higher-order corner states protected by chiral symmetry at zero energy and the higher-order topological phase can be described by quadrupole chiral number $N_{xy}$. Just like the case in 1D, it is calculated by $U_{\sigma}$ matrices, whose unitarity makes sure that $N_{xy}$ only takes integer values.

\begin{figure}[t]
	\centering
	\includegraphics[width=1\linewidth]{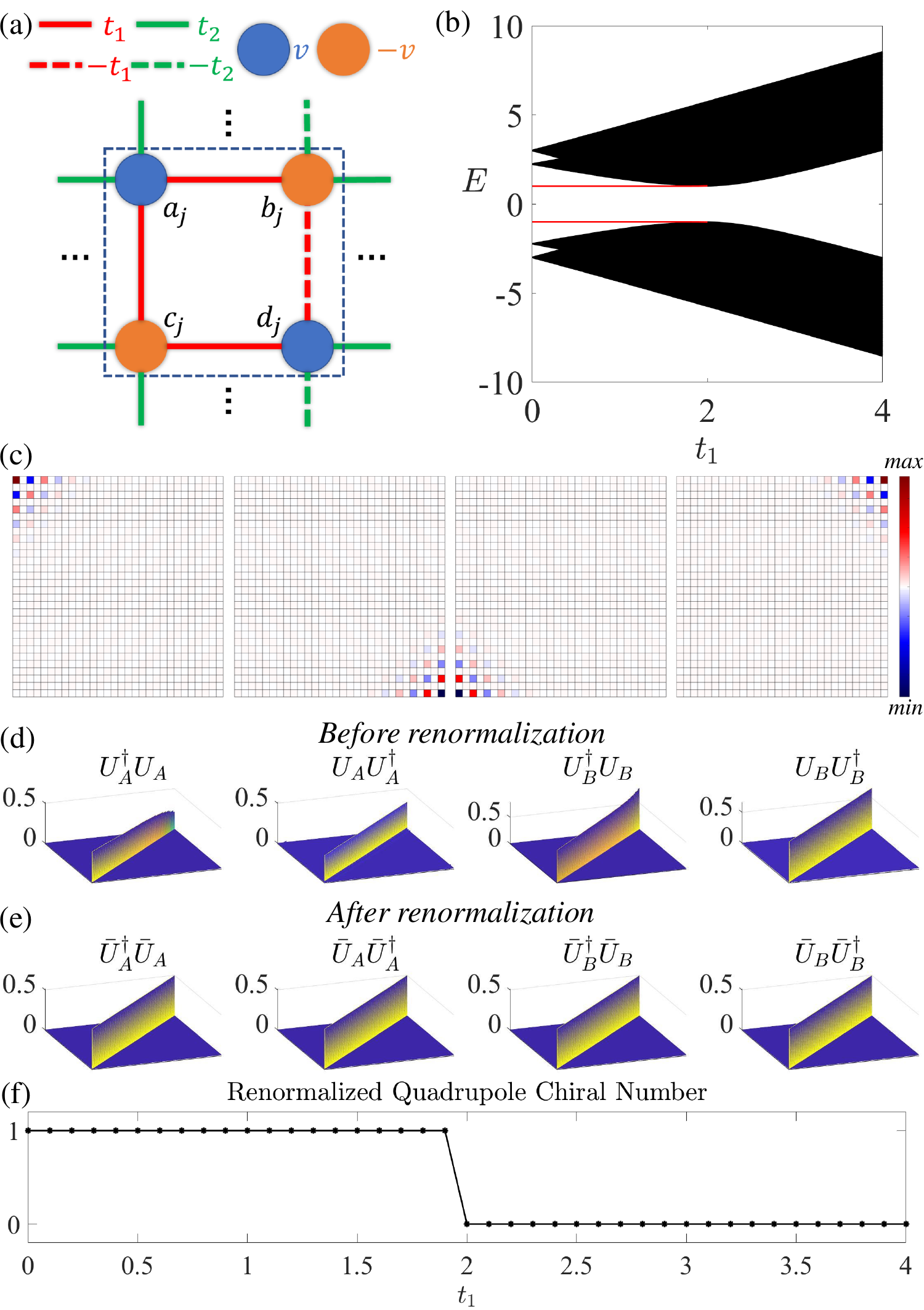}
	\caption{(a) Schematic of 2D Rice-Mele model. The unit cell is marked by dashed box. (b) Band structure versus $t_1$ with $t_2=2$, $v=1$. (c) The spatial distribution of the four corner states of 2D Rice-Mele model with $t_1=1$, $t_2=2$, $v=1$, exhibiting characteristics of chiral corner states. (d-e) The distributions of the four product matrices $U_AU_A^{\dagger}$, $U_A^{\dagger}U_A$, $U_BU_B^{\dagger}$, $U_B^{\dagger}U_B$ calculated with or without renormalization, respectively. (f) The renormalized qradrupole chiral number calculated based on the renormalized $\bar{U}_A$, $\bar{U}_B$, exhibiting precise bulk-edge correspondence. } \label{fig:3}
\end{figure}

Based on our isomorphic mapping method, we will show that the nontrivial topology in the half-filling gap for the asymmetric 2D Rice-Mele model can also be reconstructed. As shown in Fig. \ref{fig:3}(a), the 2D Rice-Mele model can be realized by assigning staggered onsite energy to the BBH model. Its Bloch Hamiltonian can be written as:
\begin{eqnarray} \label{eq21}
	\mathcal{H}_{2D-RM}(\bm{k})=\begin{bmatrix}
		vI_2 & M(\bm{k})\\
		M(\bm{k})^{\dagger} & -vI_2
	\end{bmatrix},
\end{eqnarray}
where
\begin{eqnarray} \label{eq22}
	M(\bm{k})=\begin{bmatrix}
		t_1+t_2e^{ik_x}&t_1+t_2e^{ik_y}\\
		-t_1-t_2e^{-ik_y}&t_1+t_2e^{-ik_x}\\
	\end{bmatrix}.
\end{eqnarray}
$\mathcal{H}_{2D-RM}(\bm{k})$ possesses a typical Rice-Mele-like form, which indicates that the 2D Rice-Mele model can be strictly mapped into the BBH model with a mapping relationship similar to Eq. \eqref{eq15}. As a result, in the topologically nontrivial region, namely $t_2>t_1$, two pairs of corner states with the energy of $\pm v$ are expected to appear, which agree with the numerical results shown in Fig. \ref{fig:3}(b). The spatial distributions of the four corner states, shown in Fig. \ref{fig:3}(c), also exhibit typical characteristics of corner states with chiral symmetry, where each corner state occupies only one of the four sublattices.

As to the higher-order topological invariant, if we follow the method in the original work \cite{PhysRevLett.128.127601}, the calculated ${N}_{xy}$ is not quantized since the original $U_{\sigma}$ matrices are not unitary, which is shown in Fig. \ref{fig:3}(d). With isomorphic mapping, the $U_\sigma$ can be replaced by unitary matrices $\bar{U}_{\sigma}$, whose column is the renormalized eigenstates of occupied bands on the $\sigma$ sublattice. The unitary properties of $\bar{U}_{\sigma}$ are shown in Fig. \ref{fig:3}(e). Therefore, we can redefine the renormalized quadrupole chiral number $\bar{N}_{xy}$ by $\bar{U}_{\sigma}$:
\begin{eqnarray} \label{eq19}
	\bar{N}_{xy}=\frac{1}{2\pi i} \mathrm{Tr} \left [ \mathrm{log}(\bar{Q}_{xy}^{A}\bar{Q}_{xy}^{B\dagger}) \right ],
\end{eqnarray}
here we have:
\begin{eqnarray} \label{eq20}
	\bar{Q}_{xy}^{\sigma}=\bar{U}_{\sigma}^{\dagger}{Q}_{xy}^{\sigma}\bar{U}_{\sigma},
\end{eqnarray}
for $\sigma=A,B$.
${Q}_{xy}^{\sigma}={\textstyle \sum_{\bm{R},\alpha \in \sigma} | \bm{R},\alpha \rangle e^{-i\frac{2\pi xy}{L_xL_y}} \langle \bm{R},\alpha |}$ is the sublattice quadrupole moment operator. As shown in Fig. \ref{fig:3}(f), the renormalized quadrupole chiral number is quantized and an accurate bulk-edge correspondence can be established.

\section{\label{sec:level1}Topology Reconstruction via Perturbation Approximation}

In this section, we would like to introduce another general method of topology reconstruction in asymmetry systems, i.e. the topology reconstruction based on the subspace from the perturbation approximation. Unlike previous research on multi-band systems whose subspaces rely on extra symmetry \cite{WOS:000570035500003,PhysRevB.106.205111,PhysRevLett.130.106301}, here the subspace is obtained through degenerate perturbation approximation which doesn't require any symmetry. As an example, the topological property of the Rice-Mele ladder in the $1$st and $3$rd gap (fractional-filling gaps) can be reconstructed in the subspace obtained by perturbation theory and further described by the renormalized real-space winding number with new effective sublattices. Therefore, the topology of such systems could be clearly shown with the condition of perturbation approximation. To our surprise, the topological edge states and the topological properties of such asymmetric systems exhibit extreme robustness which can go far beyond \emph{the weak perturbation condition}. This reveals the possibility that many phenomena in asymmetry systems, which are generally ignored by the topological study community, might still be dominated by deeper nontrivial topology.

\begin{figure}[t]
	\centering
	\includegraphics[width=1\linewidth]{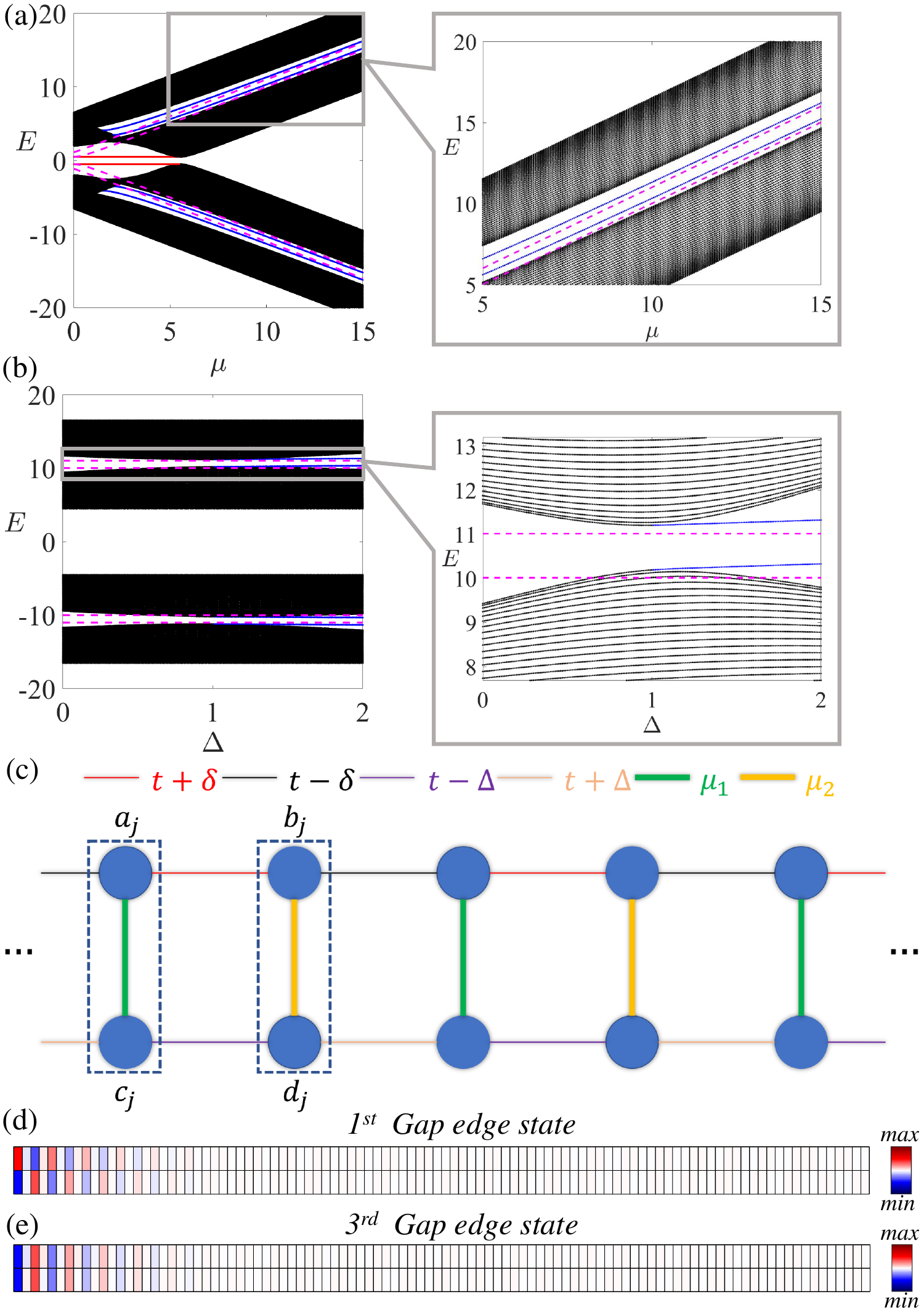}
	\caption{(a) Band structure versus $\mu$ with $t=3$, $\delta=1$, $\Delta=2$, $v=0.5$, $\mu_1=\mu$, $\mu_2=\mu+1$. (b) Band structure versus $\Delta$ with $t=3$, $\delta=1$, $\mu_1=10$, $\mu_2=11$, $v=0.5$. In both (a) and (b), $\pm\sqrt{\mu_1^2+v^2}$, $\pm\sqrt{\mu_2^2+v^2}$ are marked by red dashed lines. (c) Schematic of Rice-Mele ladder after isomorphic mapping in the strong coupling limit. The dimers in $y$ direction are marked by dashed boxes. (d-e) The renormalized spatial distribution of the edge state in (d) the 1st gap; (e) the 3rd gap with $t=3$, $\delta=1$, $\Delta=2$, $\mu_1=10$, $\mu_2=11$, $v=0.5$.} \label{fig:4}
\end{figure}

Next we will show the topology reconstruction process for the Rice-Mele ladder in the $1$st and $3$rd gaps (fractional-filling gaps) step by step. First, we need to observe the band-gap structure to choose a proper range for the perturbation approximation. The band structure of Rice-Mele ladder versus $\mu$ is shown in Fig. \ref{fig:4}(a) with $\mu_1=\mu$, $\mu_2=\mu+1$. Four non-zero edge states marked by blue lines can be observed with OBC. As the coupling terms $\mu_1$, $\mu_2$ increase, the energies of four edge states approach the asymptotes $E=\pm\sqrt{\mu_1^2+v^2}$ and $E=\pm\sqrt{\mu_2^2+v^2}$ which are marked by red dashed lines in Fig. \ref{fig:4}(a) and (b), respectively. Furthermore, as shown in Fig. \ref{fig:4}(a), in the weak coupling limit (i.e. $\mu_1,\mu_2\rightarrow 0$) both the $1$st and $3$rd gap vanish and the edge states merge into the bulk band. The band-gap structure versus $\Delta$ is shown in Fig. \ref{fig:4}(b), which exhibits a typical Rice-Mele-like topological phase transition without gap closing and reopening. The critical point can be found at $\delta=\Delta$, which means that the inter-(intra-)cell hopping of the upper chain is exactly the same as the intra-(inter-)cell hopping of the lower chain at the critical point.

With these observed properties of edge-states in mind, it is easy to see that the strong coupling limit, i.e. $\mu_1$, $\mu_2 \gg t$, $\delta$, $\Delta$, serves as a good example for our perturbation approximation to reconstruct the topology in the fractional-filling gaps. So, we will focus on the cases with strong coupling limit.

Before performing any further analysis, we would like to map the Hamiltonian of Rice-Mele ladder $\mathcal{H}_{RM}$ with the form of Eq. \eqref{eq7} into Eq. \eqref{eq13} and also map the wave functions and the eigen-energy based on Eq. \eqref{eq15}, which are referred to as isomorphic mapping in the last section. At here, we emphasize that the method of topology reconstruction by perturbation approximation does \emph{not} rely on the isomorphic mapping. Actually we can go directly to the perturbation approximation without the step of isomorphic mapping. The reason for this step is that it can provide more accurate results since it ensures that the band structure is symmetric to $E=0$, which is an obvious feature of the Rice-Mele ladder. In Appendix \ref{C}, we would like to exhibit another example of the 1D ladder model whose topological properties in the fractional-filling gaps can be directly reconstructed by perturbation approximation, without isomorphic mapping at all.

In the strong coupling limit $\mu_1$, $\mu_2 \gg t$, $\delta$, $\Delta$, we can suppose that two atoms in $y$ direction form a dimer and all other hopping terms in $x$ direction are perturbations. After dimerization, the schematic model is exhibited in Fig. \ref{fig:4}(c), in which two dimerized atoms are enclosed by the dashed lines. Two kinds of dimers can be found with the Hamiltonian as:
\begin{eqnarray} \label{eq23}
	\mathcal{H}_{dimer}^{A}=
	\begin{bmatrix}
		0 & \mu_1\\
		\mu_1 & 0
	\end{bmatrix}, \mathcal{H}_{dimer}^{B}=
	\begin{bmatrix}
		0 & \mu_2\\
		\mu_2 & 0
	\end{bmatrix}.
\end{eqnarray}
The eigen-energies and eigenstates of dimers can be expressed as:
\begin{eqnarray} \label{eq24}
    \left\{\begin{matrix}
		E_{1i}^A = -\mu_1, \ \psi_{1i}^A = \frac{1}{\sqrt{2}}(|i,a\rangle - |i,c\rangle)
        \\
        E_{2i}^A = \mu_1, \ \psi_{2i}^A = \frac{1}{\sqrt{2}}(|i,a\rangle + |i,c\rangle)
        \\
		E_{1i}^B = -\mu_2, \ \psi_{1i}^B = \frac{1}{\sqrt{2}}(|i,b\rangle - |i,d\rangle)
        \\
        E_{2i}^B = \mu_2, \ \psi_{2i}^B = \frac{1}{\sqrt{2}}(|i,b\rangle + |i,d\rangle)
	\end{matrix}\right. .
\end{eqnarray}
Taking the eigenstates $\{\psi_{11}^A,...,\psi_{1N}^A,\psi_{11}^B,...,\psi_{1N}^B,\psi_{21}^A,...,$ $\psi_{2N}^A,\psi_{21}^B,...,\psi_{2N}^B\}$ as a set of new basis functions, then the Hamiltonian can be rewritten as:
\begin{eqnarray} \label{eq25}
	\tilde{\mathcal{H}}=\begin{bmatrix}
		\mathcal{H}^{11} & \mathcal{H}^{12} \\
		\mathcal{H}^{21} & \mathcal{H}^{22}
	\end{bmatrix},
\end{eqnarray}
where each $\mathcal{H}^{ij}$ stands for a $(2N) \times (2N)$ matrix:
\begin{eqnarray}
	\mathcal{H}^{11}=\begin{bmatrix} \label{eq26}
        -\mu_1 I_N & h_{chain} \\
        h_{chain}^{\dagger} & -\mu_2 I_N
	\end{bmatrix},\\
    \mathcal{H}^{22}=\begin{bmatrix} \label{eq27}
        \mu_1 I_N & h_{chain} \\
        h_{chain}^{\dagger} & \mu_2 I_N
	\end{bmatrix}, \\
    \mathcal{H}^{12}= \mathcal{H}^{21 \dagger}=\begin{bmatrix} \label{eq28}
        0 & h_{couple} \\
        h_{couple}^{\dagger} & 0
	\end{bmatrix},
\end{eqnarray}
with
\begin{eqnarray}
	h_{chain}=\begin{bmatrix} \label{eq29}
        t_1 & & & & & \\
        t_2 & t_1 & & & & \\
         & t_2 & t_1 & & & \\
         & & \ddots & \ddots  & & \\
         &  &  &  t_2 & t_1
	\end{bmatrix}_{N \times N},\\
    h_{couple}=\begin{bmatrix} \label{eq30}
        t' & & & & & \\
        -t' & t' & & & & \\
         & -t' & t' & & & \\
         & & \ddots & \ddots  & & \\
         &  &  &  -t' & t'
	\end{bmatrix}_{N \times N}.
\end{eqnarray}
Here $t_1=t+\frac{1}{2}(\delta-\Delta)$, $t_2=t-\frac{1}{2}(\delta-\Delta)$, $t'=\delta+\Delta$. When $\mu_1$, $\mu_2$ are relatively close, i.e. $\mu_1$, $\mu_2 \gg |\mu_1-\mu_2|$, $\tilde{\mathcal{H}}$ can be brought into two parts, i.e. $\tilde{\mathcal{H}}=\mathcal{H}_0+\mathcal{H}'$, where $\mathcal{H}_0$ stands for the unperturbed Hamiltonian while $\mathcal{H}'$ is treated as the perturbation. Here $\mathcal{H}_0$ is defined as:
\begin{eqnarray} \label{eq31}
	\mathcal{H}_0=\begin{bmatrix}
		-\frac{1}{2}(\mu_1+\mu_2)I_{2N} & 0 \\
		0 & \frac{1}{2}(\mu_1+\mu_2)I_{2N}
	\end{bmatrix}.
\end{eqnarray}
As $\mathcal{H}_0$ is already a diagonal matrix, it generates two sets of $2N$-fold degenerate states $\{\psi_{1i}^\sigma\},\{\psi_{2i}^\sigma\},i=1,...,N,\sigma=A,B$ with the corresponding zero-order energies $E^{(0)}=\pm \frac{1}{2}(\mu_1+\mu_2)$. Then $\mathcal{H}'$ can be written as:
\begin{eqnarray} \label{eq32}
	\mathcal{H}'=\begin{bmatrix}
		\mathcal{H}^{11'} & \mathcal{H}^{12} \\
		\mathcal{H}^{21} & \mathcal{H}^{22'}
	\end{bmatrix},
\end{eqnarray}
with
\begin{eqnarray}
	\mathcal{H}^{11'}=\begin{bmatrix} \label{eq33}
        \frac{1}{2}(\mu_2-\mu_1) I_N & h_{chain} \\
        h_{chain}^{\dagger} & \frac{1}{2}(\mu_1-\mu_2) I_N
	\end{bmatrix},\\
    \mathcal{H}^{22'}=\begin{bmatrix} \label{eq34}
        \frac{1}{2}(\mu_1-\mu_2) I_N & h_{chain} \\
        h_{chain}^{\dagger} & \frac{1}{2}(\mu_2-\mu_1) I_N
	\end{bmatrix},
\end{eqnarray}
where $h_{chain}$ is defined in Eq. \eqref{eq29}. Obviously, $\mathcal{H}'$ satisfies the perturbation condition since $E^{(0)} \gg \mathcal{H}'_{mn}$ is valid. Therefore, the solutions of $\tilde{\mathcal{H}}$ can be approached through perturbation approximation.

Here, we would focus on the degenerate perturbation to obtain the zero-order wave function $\varphi^{(0)}$ and the first-order energy correction $E^{(1)}$. The principle of degenerate perturbation theory is to diagonalize the perturbation Hamiltonian in the degenerate subspace, i.e. to diagonalize $\mathcal{H}^{11'}$ and $\mathcal{H}^{22'}$ defined in Eq. \eqref{eq33} and Eq. \eqref{eq34}, respectively. One may notice that these two subspace perturbation Hamiltonians are exactly the Hamiltonians of the Rice-Mele model with the effective onsite energy $v_{eff}= \pm \frac{1}{2}(\mu_1-\mu_2)$, the intra-cell hopping $t_1$ and the inter-cell hopping $t_2$. \emph{Therefore, $\varphi^{(0)}$ and $E^{(1)}$ are just the eigenstate and eigen-energy of the Rice-Mele model. In other words, the topological properties around $\pm \frac{1}{2}(\mu_1+\mu_2)$, i.e. the $1$st and $3$rd gap, are dominated by Rice-Mele-like topology, which is well studied in the last section.} For example, when $t_2>t_1$, i.e. $\Delta>\delta$, two pairs of edge states are expected to emerge with $E^{(1)}=\pm \frac{1}{2}(\mu_1-\mu_2)$, respectively. To be precise, the eigen-energies of these four edge states up to the first-order correction are given as $E^{(0)}+E^{(1)}=\pm \mu_1, \pm \mu_2$. Taking mapping relation Eq. \eqref{eq15} into consideration, $E^{(0)}+E^{(1)}$ can be mapped back to the physical values as $\pm\sqrt{\mu_1^2+v^2},\pm\sqrt{\mu_2^2+v^2}$ for $\mathcal{H}_{RM}$, which are exactly same as the asymptotes marked by red dashed lines in Fig. \ref{fig:4}(a-b). The perturbation theory is further confirmed by the renormalized spatial distribution of the edge states shown in Fig. \ref{fig:4}(d-e), which exhibit typical characteristics of Rice-Mele-like edge states defined on the degenerate subspaces $\{\psi_{1i}^\sigma\},\{\psi_{2i}^\sigma\}$, respectively.

We note that the coupling between two subspaces caused by $\mathcal{H}^{12}$ and $\mathcal{H}^{21}$ is negligible to the first order correction of eigen-energy, since the non-degenerate perturbation can only generate higher-order corrections which are rather minor compared with the dominating $\varphi^{(0)}$ and $E^{(1)}$ discussed in this section.

\begin{figure}
	\centering
	\includegraphics[width=1\linewidth]{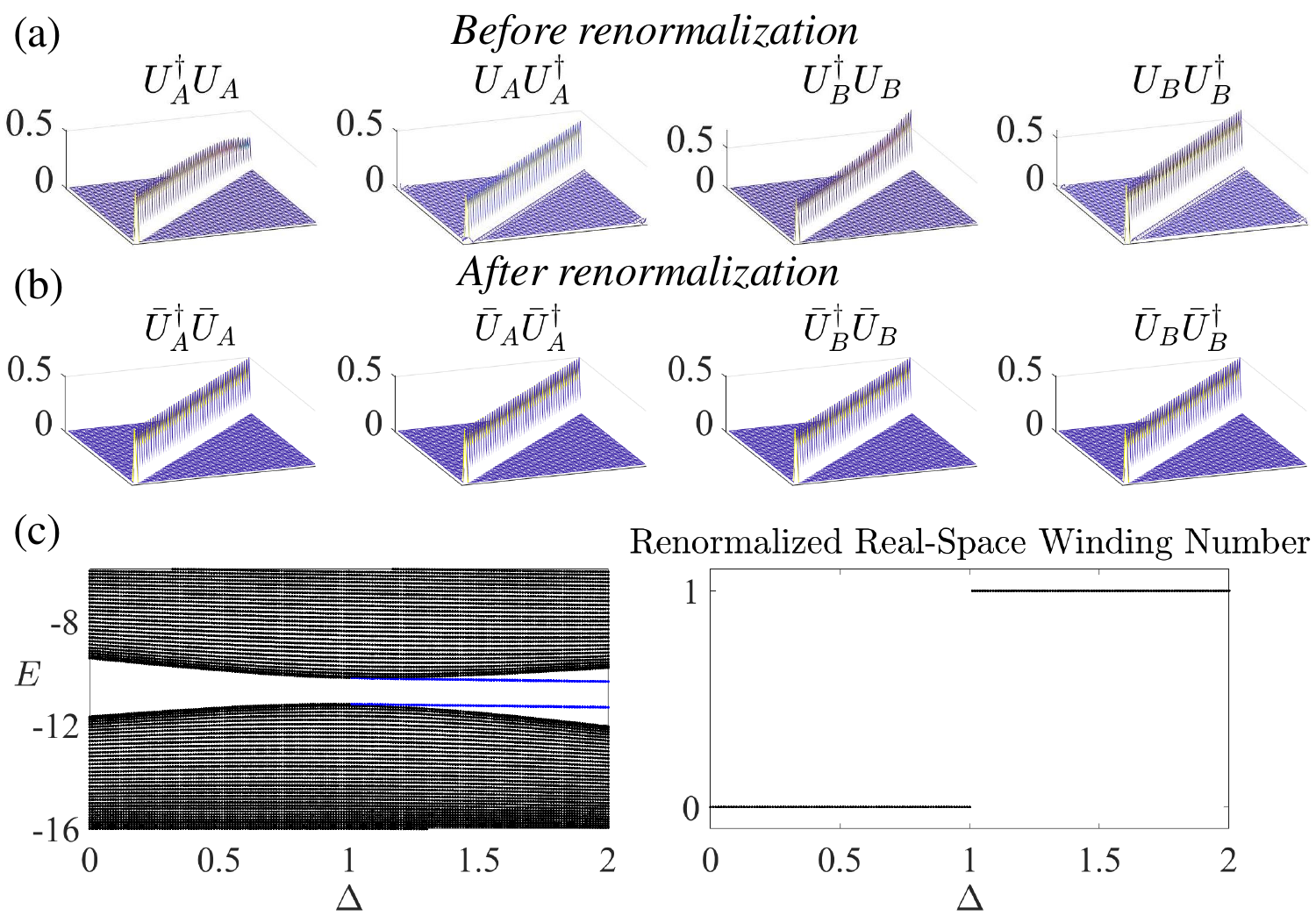}
	\caption{(a-b) Four product matrices calculated based on a redefined sublattice with or without renormalization, respectively. The parameters are set to $t=3$, $\delta=1$, $\Delta=2$, $v=0.5$, $\mu_1=10$, $\mu_2=11$. (c) The renoramlized real-space winding number exhibits precise bulk-edge correspondence with the edge states at $\frac{1}{4}$ filling. The case at $\frac{3}{4}$ filling should be the same since the band structure is symmetric to $E=0$.} \label{fig:5}
\end{figure}

Next, we would introduce the topological invariant to describe the subspace topology reconstructed above by perturbation approximation. As both subspace Hamiltonians $\mathcal{H}^{11'}$, $\mathcal{H}^{22'}$ possess a Rice-Mele-like form, it is straightforward to take renormalized real-space winding number as the candidate. Different from the case in the last section where the whole Hilbert space is under consideration, here the subspace constructed by $\{\psi_{ji}^A\},\{\psi_{ji}^B\}$ for perturbation approximation is of interest. In this sense, sublattices $A$ and $B$ should be redefined as $\{\psi_{1i}^A\}$ and $\{\psi_{1i}^B\}$ at $\frac{1}{4}$ filling and $\{\psi_{2i}^A\},\{\psi_{2i}^B\}$ at $\frac{3}{4}$ filling, respectively. With these considerations, for the topology in the $1$st gap, the $U_\sigma$ matrix can be obtained as:
\begin{eqnarray} \label{eq35}
    \left\{\begin{matrix}
        U_{A(nm)} = \langle \bar{\varphi}_m | \psi^A_{1n} \rangle
    \\
        U_{B(nm)} = \langle \bar{\varphi}_m | \psi^B_{1n} \rangle
    \end{matrix}\right. ,
\end{eqnarray}
where $n=1,...,N$, $\bar{\varphi}_m$ stands for the $m$th renormalized eigenstate below the $1$st gap, $m=1,...,N$. When parameters are set to $\mu_1=10$, $\mu_2=11$, $t=3$, $\delta=1$, $\Delta=2$, $v=0.5$, $U_AU_A^{\dagger}$, $U_A^{\dagger}U_A$, $U_BU_B^{\dagger}$, $U_B^{\dagger}U_B$ calculated based on Eq. \eqref{eq35} are shown in Fig. \ref{fig:5}(a). In Fig. \ref{fig:5}(a), the four product matrices exhibit typical characteristics of Rice-Mele-like system, confirming the validity of our method. The unitarity of $U_A$ and $U_B$ can be restored through the renormalization of Eq. \eqref{eq18}, which are shown in Fig. \ref{fig:5}(b). The calculated renormalized real-space winding number shown in Fig. \ref{fig:5}(c) exhibits a unit jump at the critical point $\Delta=\delta$, indicating the bulk-edge correspondence has been established. It is worth noting that the renormalization needs to be done twice here, where the first one maps $\mathcal{H}_{RM}$ into its off-diagonal counterpart and the second one maps the subspace $\mathcal{H}^{11'}$ into its off-diagonal counterpart.

\begin{figure}[t]
	\centering
	\includegraphics[width=1\linewidth]{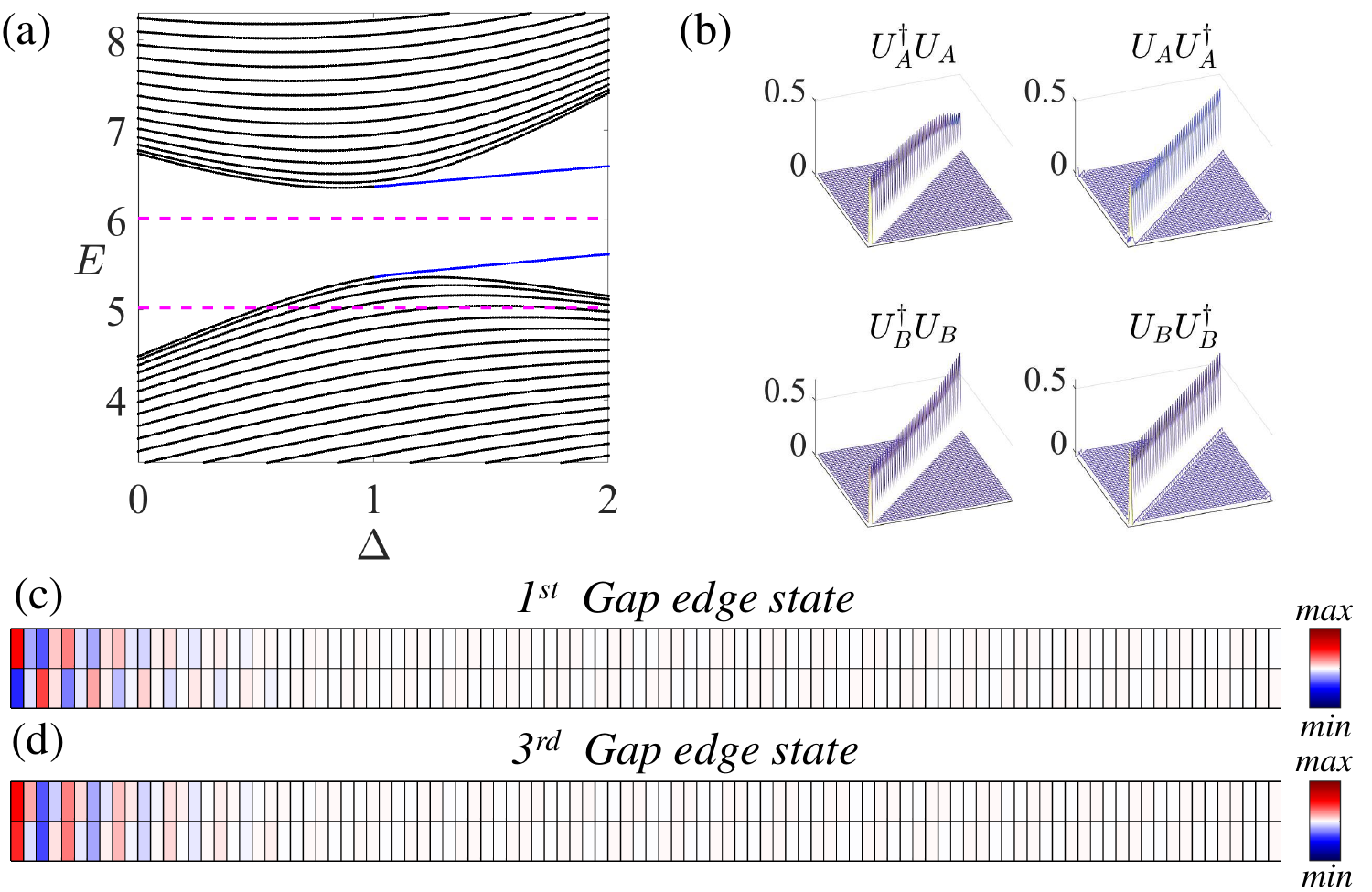}
	\caption{ The subspace topology of Rice-Mele ladder beyond the perturbation region with $\mu_1=5$, $\mu_2=6$, $t=3$. (a) The band structure versus $\Delta$ with $\delta=1$, $v=0.5$ around the $3$rd gap. The $1$st gap shares the same structure. (b) Four product matrices calculated based on a redefined sublattice without renormalization, also maintaining its Rice-Mele-like feature. (c-d) The spatial distribution of the edge states in (c) the $1$st gap; (d) the $3$rd gap, still exhibiting characteristics of Rice-Mele-like edge states defined on the subspaces despite a stronger coupling from the other subspace. } \label{fig:6}
\end{figure}

Similarly, for the topology in the $3$rd gap, the $U_\sigma$ matrix is given by:
\begin{eqnarray} \label{eq36}
	\left\{\begin{matrix}
		U_{A(nm)} = \langle \bar{\varphi}_m | \psi^A_{2n} \rangle
		\\
		U_{B(nm)} = \langle \bar{\varphi}_m | \psi^B_{2n} \rangle
	\end{matrix}\right. .
\end{eqnarray}
Here $n=1,...,N$. However, if we directly take all the eigenstates below the $3$rd gap, i.e. $m=1,...,3N$, Eq. \eqref{eq6} would never be satisfied since both $U_A$ and $U_B$ are no longer square matrices. An alternative way is to consider the general fact that a fully occupied system is topologically trivial. That is to say, the occupied band(s) share the same topology as the unoccupied one(s). Therefore, the $U_\sigma$ matrix for the $3$rd gap can be constructed with the states from $3N+1$ to $4N$ and the topology of the $3$rd gap can also be described by renormalized real-space winding number.

\emph{Interestingly, the subspace topology can go far beyond the proper range of the perturbation approximation.} Fig. \ref{fig:6} exhibits the case when $\mu_1$, $\mu_2$ are comparable with $t$, $\delta$, $\Delta$. From the band structure shown in Fig. \ref{fig:6}(a), the edge states marked by blue lines still appear in the $1$st and $3$rd gap despite an obvious shift from the asymptotes $\pm\sqrt{\mu_1^2+v^2}$, $\pm\sqrt{\mu_2^2+v^2}$ marked by red dashed lines. As shown in Fig. \ref{fig:6}(b-d), the four product matrices calculated based on Eq. \eqref{eq35} and the renormalized spatial distribution also maintain their subspace Rice-Mele-like characteristics. This phenomenon can be understood as the interplay between perturbation and topology. On one hand, we have clearly shown that the subspace topology is reconstructed based on perturbation approximation. On the other hand, the topology itself, which is the global geometric property of the bands, should be robust against the continuous change of the system parameters. That is to say, in the topologically non-trivial region, namely $\Delta>\delta$, the edge states in the fractional-filling gaps are rather robust as long as $\mu_1$, $\mu_2$ are strong enough to keep the $1$st and $3$rd gap open, even if the perturbation condition is no longer satisfied.

The idea introduced in this section is one of the main conclusions in this work. First, we find that perturbation theory can greatly extend the category of topology study. Taking the merit of degenerate perturbation, the whole Hilbert space can be divided into degenerate subspaces where the topological properties might be revealed, even in asymmetric systems that were previously thought to be topologically irrelevant. Second, some new topics can immediately be raised when the perturbation approximation is introduced, such as "Where is the boundary of the perturbation-based topological nontriviality?" and "Is there any topological structure from the higher-order perturbation?", etc.

Furthermore, as shown in the excellent works \cite{PhysRevB.98.024205,PhysRevB.90.014505,PhysRevResearch.3.013122}, the Hamiltonian of Kitaev chain in the Majorana representation mimics the one of SSH ladder (i.e. Rice-Mele ladder with $v=0$), indicating the topological phases of SSH ladder can be introduced into superconducting systems. However, constrained by chiral symmetry, previous works have always focused on the topological properties of zero edge states in the half-filling gap. Taking advantage of the topology reconstruction method introduced in this section, we unveil the new Rice-Mele-like fermionic topology in the dimerized Kitaev chain whose properties are totally different from the traditional symmetry-based topology. The detailed derivation is presented in Appendix \ref{D}.

\section{\label{sec:level1}OBSERVATION OF TOPOLOGICAL EDGE STATES FROM TOPOLOGY RECONSTRUCTION IN PHOTONIC CRYSTALS}
In this section we would like to introduce both methods of the topology reconstruction into dielectric PhC systems.
This work can be easily done since TBM can be realized by artificial atoms in dielectric 2D PhCs \cite{PhysRevB.97.035442,PhysRevLett.122.233903,PhysRevLett.122.233902}.

First, we set up the model of photonic Rice-Mele ladder and construct the photonic TBM theory for our model. Then both methods of topology reconstruction for asymmetric photonic systems, the isomorphic mapping method and the perturbation approximation method, are demonstrated. We also show that the topological invariant obtained by our theory from photonic tight-binding eigenstate, namely the renormalized real-space winding number, matches precisely with the numerical results obtained from the finite element method (FEM). At last, in the laboratory, the topological edge states from both kinds of topology reconstruction for asymmetric systems are observed through the experiments, which agree well with our theoretical prediction. 	

\subsection{\label{sec:level2}Photonic Rice-Mele Ladder}

\begin{figure}[t]
	\centering
	\includegraphics[width=1\linewidth]{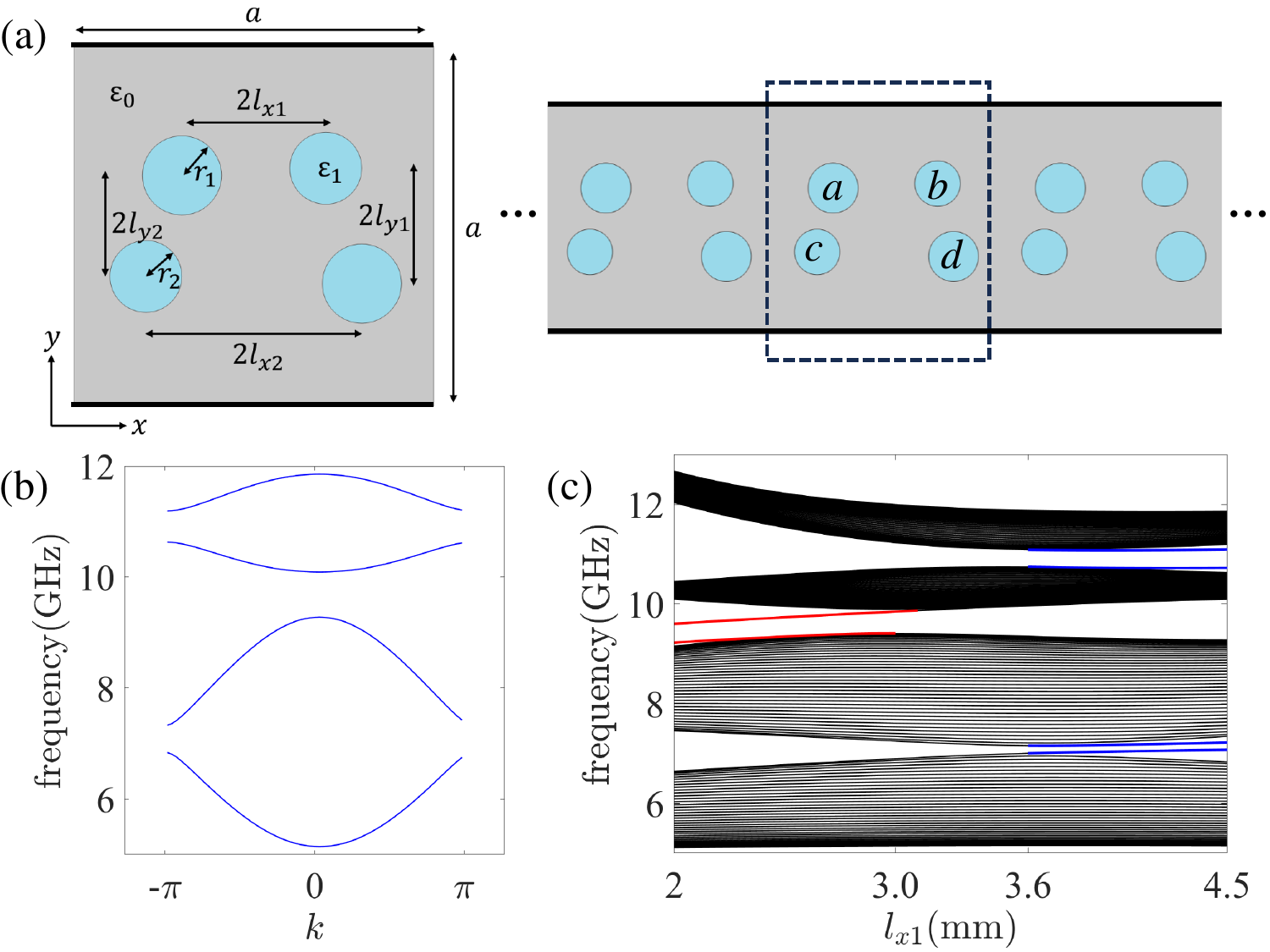}
	\caption{(a) Schematic of photonic Rice-Mele ladder. In this work, we set $l_{x1}$ as the tuning parameter. Other parameters are set to $a=18$mm, $r_1=2$mm, $r_2=1.8$mm, $l_{x2}=5.4$mm, $l_{y1}=2.5$mm, $l_{y2}=2.9$mm, $\varepsilon_0=1$, $\varepsilon_1=7.9$. PEC boundary in $y$ direction is marked by thick black line. The four rods are marked by $a$, $b$, $c$, $d$, respectively. (b) The lowest four TM bands of photonic Rice-Mele ladder with $l_{x1}=4.5$mm. (c) The band structure versus $l_{x1}$ for 40-unit-cell photonic Rice-Mele ladder with PEC boundary in both $x$ and $y$ directions. Two types of toplogical edge states are marked by blue lines and red lines, respectively. } \label{fig:7}
\end{figure}

The model of a PhC unit cell of the photonic Rice-Mele ladder is shown in Fig. \ref{fig:7}(a), which is periodic in $x$ direction while both boundaries in $y$ direction are taken as perfect electric conductor (PEC). Each cell consists of four dielectric rods with $\varepsilon_1$ marked by $a$, $b$, $c$, $d$, which are further divided into two pairs diagonally with radius $r_1$ and $r_2$, respectively. We take the $x$-direction spacing between upper rods $l_{x1}$ as the tuning parameter to control the topological phase transition. Other parameters are set to $a=18$mm, $r_1=2$mm, $r_2=1.8$mm, $l_{x2}=5.4$mm, $l_{y1}=2.5$mm, $l_{y2}=2.9$mm, $\varepsilon_0=1$, $\varepsilon_1=7.9$. Obviously, the model is spatially asymmetric in every aspect. Based on Mie resonance theory \cite{PhysRevApplied.18.064089}, the corresponding relation between photonic model and TBM can be established as the following laws. First, the Mie resonant frequency of dielectric rod, which can be adjusted by altering the radius $r_1$ or $r_2$, mimics the onsite energy in TBM. Second, the coupling strength between different rods, which can be adjusted by altering the spacing between rods, mimics the hopping terms in TBM. In our model, the difference between $r_1$ and $r_2$ is relatively minor in order to suppress the next-nearest-neighbor (NNN) coupling, which is neglected in our theory.

For the convenience of experimental realization, here we focus on the TM ($E_z$ polarization) modes of our PhC structure. The four lowest TM bands with PBC are shown in Fig. \ref{fig:7}(b). As we expect, the band-gap structure of PhC is similar to the Rice-Mele ladder investigated above. In Fig. \ref{fig:7}(c), we also calculate the band structure of 40-unit-cell PhC with PEC boundary in both $x$ and $y$ directions versus the tuning parameter $l_{x1}$. When $l_{x1}<3.0$mm, edge states marked by red lines can be observed in the $2$nd gap. When $l_{x1}>3.6$mm, edge states marked by blue lines can be observed in the $1$st and $3$rd gap. We will see that all of these edge states are topologically protected and the corresponding topological phases can be described by the topological invariant introduced in the next subsection.

\subsection{\label{sec:level2} The Photonic Renormalized Real-Space Winding Number and Topological Phases}
In this subsection, we would derive the method to calculate the photonic renormalized real-space winding number, from which the topological phases of the photonic Rice-Mele ladder can be signified clearly.

\begin{figure}[t]
	\centering
	\includegraphics[width=1\linewidth]{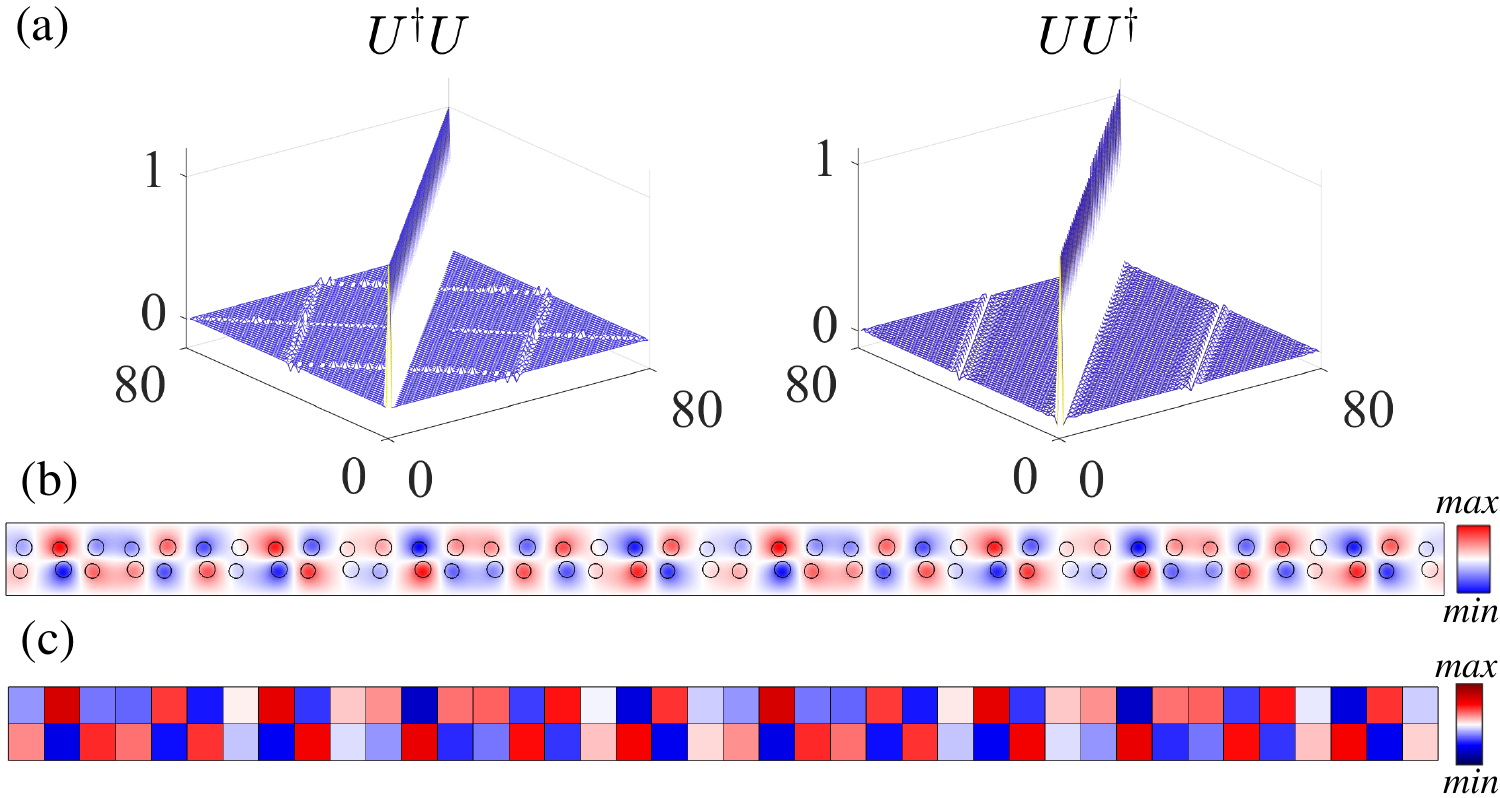}
	\caption{(a) The distributions of product matrices $UU^{\dagger}$, $U^{\dagger}U$. The $i$th column of the $U$ matrix is the $i$th tight-binding PhC eigenstate. (b-c) The spatial distribution of a PhC eigenstate obtained by (b) FEM and (c) Eq. \eqref{eq37}.} \label{fig:8}
\end{figure}

To calculate the renormalized real-space winding number of the photonic systems, we must first obtain the tight-binding version of our PhC eigenstate, i.e. $\psi={\textstyle \sum_{j,\alpha}C_{j,\alpha}|j,\alpha \rangle }$ where $j$ is the index of unit cell, $\alpha=a$, $b$, $c$, $d$ is the index of sublattice and $C_{j,\alpha}$ is the effective probability amplitude of site $|j,\alpha \rangle$. Generally speaking, this can be done by expanding the continuous eigenstate $E(\bm{x})$ obtained by FEM in the maximally localized Wannier functions (MLWF)\cite{PhysRevB.97.035442}. However, this method can be rather complicated and inconvenient. In this work, we develop an approximate method to obtain $C_{j,\alpha}$ in a much easier way. According to Mie theory, the origin of lower-order PhC bands can be understood as the coupling between the first-order Mie resonant states of the dielectric rods, which is mainly concentrated on the rod. With all these basic observations and the original idea of TBM that the electronic wavefunction could be represented by a complex amplitude at the position of nuclear, we can suppose that the TBM of our lowest four bands could be similarly defined. First, we can assume a $\theta$-function:
\begin{eqnarray} \label{eq38}
	\theta_{j,\alpha}(\bm{x})=\left\{\begin{matrix}
		\frac{1}{\sqrt{S_\alpha \varepsilon_1}},\ \text{inside the rod};
		\\
		0,\ \text{outside the rod}.
	\end{matrix}\right.
\end{eqnarray}
Here $S_\alpha$ stands for the cross-sectional area of the $\alpha$ rod. Then, we can obtain the approximate TBM of PhC for the lowest four bands as:
\begin{eqnarray} \label{eq37}
    C_{j,\alpha}^i = \int \theta_{j,\alpha}^*(\bm{x})E^{i}(\bm{x})\varepsilon(\bm{x}) d\bm{x},
\end{eqnarray}
where $E^{i}(\bm{x})$ is the $i$th eigenstate of $N$-unit-cell PhC with PBC \cite{PhysRevB.103.224208} obtained by FEM, $i=1,...,4N$ ($N$ eigenstates per band).

To check the property of our approximate TBM of PhC, we exhibit the unitarity of $U$ matrix whose $i$th column is the $i$th tight-binding PhC eigenstate and the results are shown in Fig. \ref{fig:8}(a). It's clear that both $UU^{\dagger}$ and $U^{\dagger}U$ are almost identity matrices except the negligible off-diagonal elements, which indicates the orthogonality and normalization properties of these approximate TBM eigenstates. Moreover, the effectiveness of our approximate method is proved from the comparison between eigenstates obtained by FEM and our approximate TBM in Fig. \ref{fig:8}(b-c). Most importantly, with the tight-binding eigenstates, we can directly calculate the renormalized real-space winding number to identify the topological phases of such asymmetric photonic system.

\begin{figure}[t]
	\centering
	\includegraphics[width=1\linewidth]{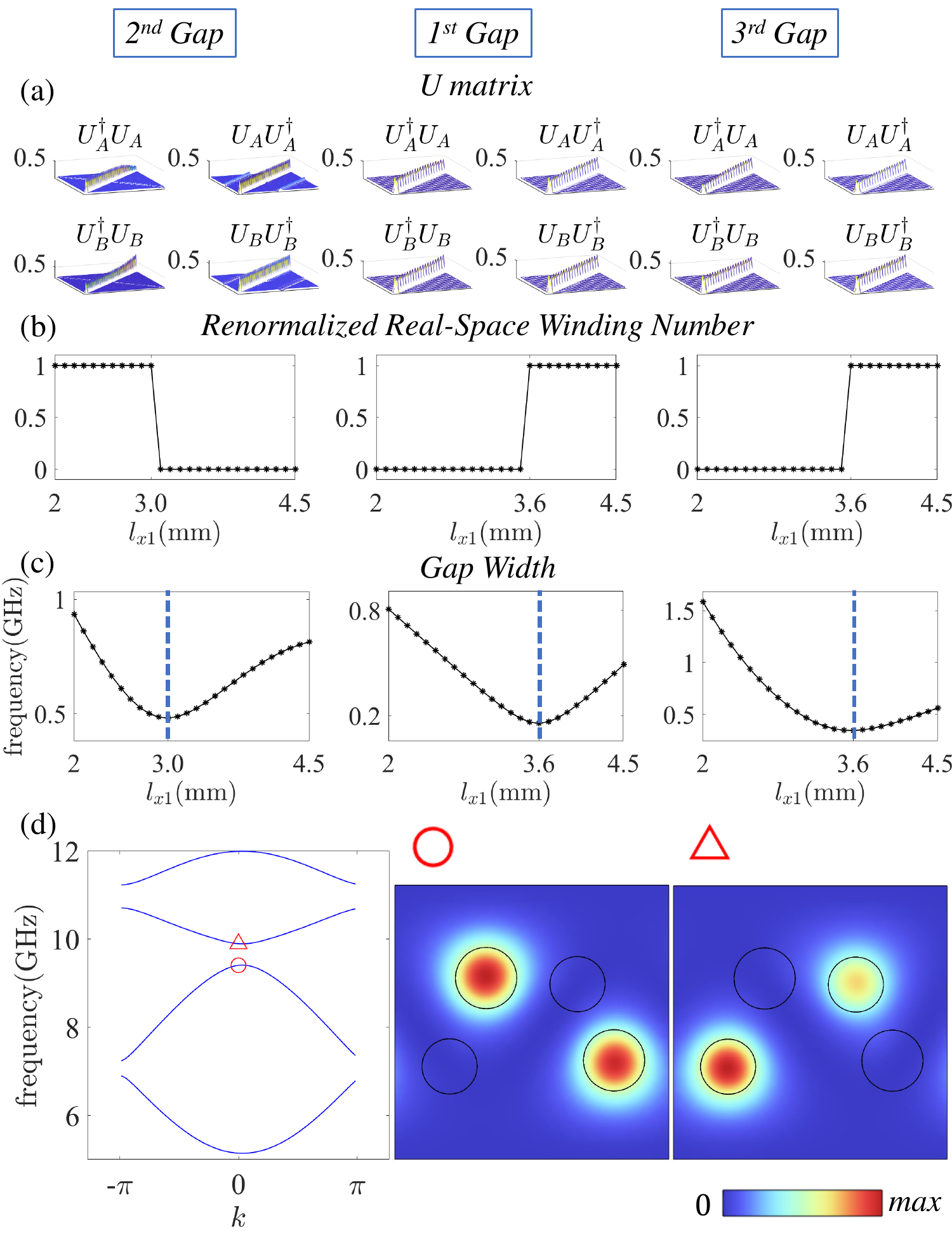}
	\caption{(a) The four product matrices $U_AU_A^{\dagger}$, $U_A^{\dagger}U_A$, $U_BU_B^{\dagger}$, $U_B^{\dagger}U_B$ calculated without renormalization for the $2$nd, $1$st, $3$rd gap, respectively. (b) The renormalized real-space winding number of the $2$nd, $1$st, $3$rd gap versus $l_{x1}$. (c) The width of the $2$nd, $1$st, $3$rd gap versus $l_{x1}$. (d) The $|E_z|^2$ distribution of the $2$nd and $3$rd Bloch eigenstates with $k=0$ at $l_{x1}=3.0$mm. Both of them are chiral eigenstates, which can be regarded as an evidence of the topological phase transition point. } \label{fig:9}
\end{figure}

First, we consider the topology of the $2$nd gap, which corresponds to the gap around zero energy for Rice-Mele ladder whose topology reconstruction method is based on isomorphic mapping which is introduced in the Section \uppercase\expandafter{\romannumeral3}. In this case, rods $a$ and $d$ are defined as sublattice $A$ while rods $b$ and $c$ as sublattice $B$. The corresponding four product matrices $U_AU_A^{\dagger}$, $U_A^{\dagger}U_A$, $U_BU_B^{\dagger}$, $U_B^{\dagger}U_B$ shown in Fig. \ref{fig:9}(a) exhibit typical characteristics of Rice-Mele-like system shown in Fig. \ref{fig:2}(b). Following the way of renormalization, we can obtain the renormalized matrices $\bar{U}_A$, $\bar{U}_B$ satisfying Eq. \eqref{eq6}, with which the renormalized real-space winding number can be calculated. Shown in the first column of Fig. \ref{fig:9}(b), the renormalized real-space winding number shows the quantized property and a unit jump from $1$ to $0$ appears at around $l_{x1}=3.0$mm, which agrees with disappearing of the edge states in Fig. \ref{fig:7}(c).

The physical reason for the topological phase transition at $l_{x1}=3.0$mm can be explained from two aspects. In terms of band structure, as shown in the first column of Fig. \ref{fig:9}(c), the gap width of the $2$nd gap achieves minimum at $l_{x1}=3.0$mm. In terms of gap-edge eigenstates, as shown in Fig. \ref{fig:9}(d), we find that both Bloch states at $k=0$ marked by triangle and circle are chiral eigenstates. These two properties of topological phase transition agree with the typical band structure characteristics of Rice-Mele-like topology discussed in Section \uppercase\expandafter{\romannumeral3}.

Second, we will discuss the case of the $1$st and $3$rd gap whose topology reconstruction is based on the degenerate perturbation approximation as introduced in Section \uppercase\expandafter{\romannumeral4}. Before the discussion, we need to note that there is a significant difference between the eigenstates of photonic systems and that of electronic systems described by Eq. \eqref{eq1} for the $1$st and the $3$rd gaps. As mentioned above, the topology of the $1$st gap in Rice-Mele ladder is dominated by the low-energy state of the dimer in $y$ direction, i.e. the $\begin{matrix}[1 &  -1] \end{matrix}^T$ state. For our PhC model discussed in this section, taking its waveguide properties into consideration, the behavior of PhC eigenstates in $y$ direction should be determined by the order of the guided modes. From this point of view, the topology of the $1$st gap in PhC is dominated by the lower-order mode, i.e. the $\begin{matrix}[1 &  1] \end{matrix}^T$ state. To be precise, in our PhC model, we need to redefine the sublattices $A$ and $B$ in Eq. \eqref{eq24}, i.e. $\{\psi_{2i}^A\}$, $\{\psi_{2i}^B\}$ for the $\frac{1}{4}$ filling gap while $\{\psi_{1i}^A\}$, $\{\psi_{1i}^B\}$ for the $\frac{3}{4}$ filling gap. We note that this phenomenon could be explained by the negative effective hopping in classical systems, which also widely exists in sonic crystals \cite{PhysRevLett.130.106301}. The corresponding $U_\sigma$ matrices and the renormalized real-space winding number are shown in the second and third columns of Fig. \ref{fig:9}(a-b). We can see that there is a unit jump from $0$ to $1$ at $l_{x1}=3.6$mm for the renormalized real-space winding number, which agrees with the appearing of edge states at the $1$st and $3$rd gap in Fig. \ref{fig:7}(c).

\begin{figure}[t]
	\centering
	\includegraphics[width=1\linewidth]{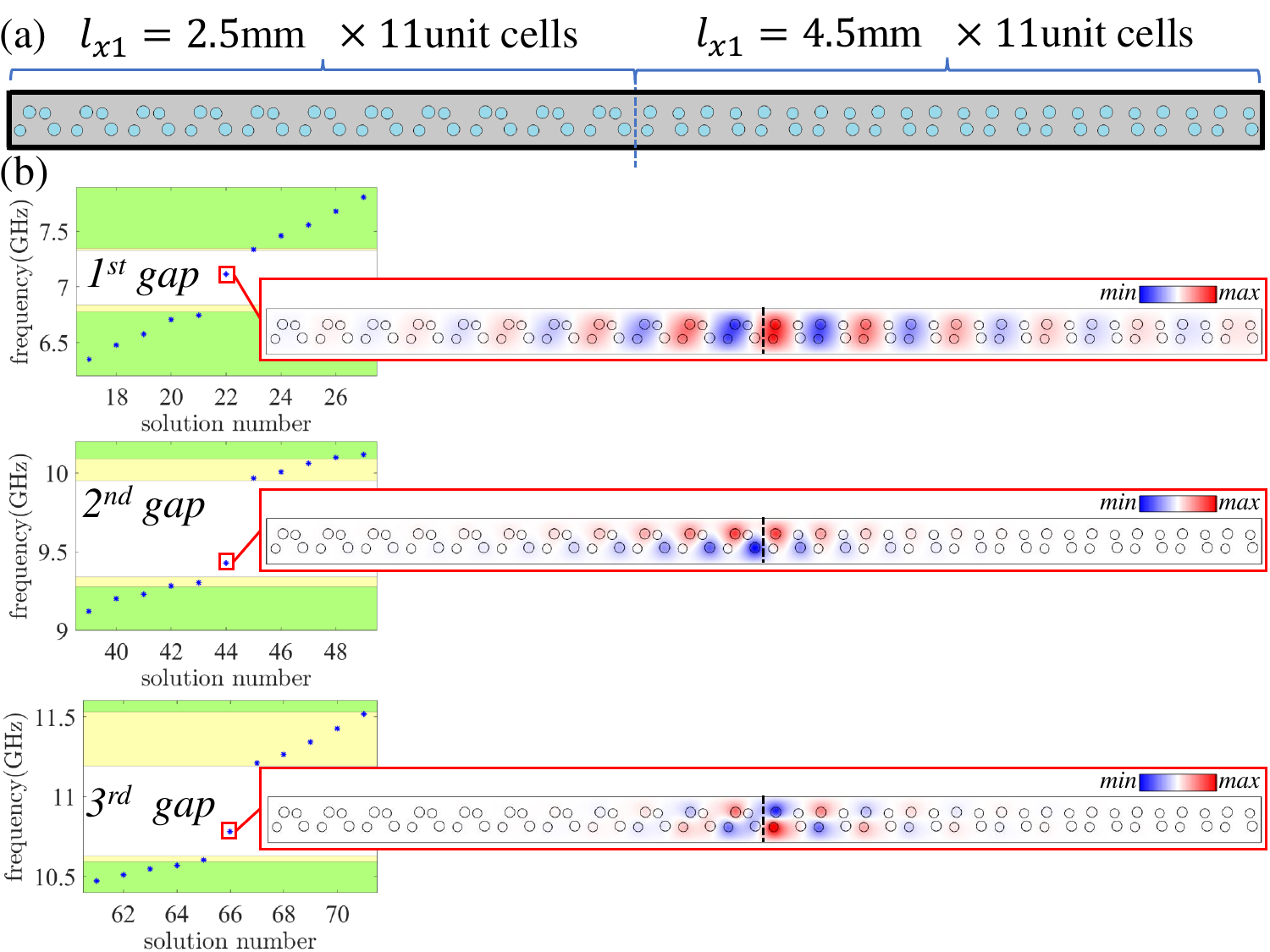}
	\caption{(a) Schematic of the bonded PhC with eleven unit cells of $l_{x1}=2.5$mm on the left and eleven unit cells of $l_{x1}=4.5$mm on the right. The whole structure is enclosed by PEC boundary marked by thick black lines. (b) Band structure of the combined PhC and the $E_z$ distribution of the three in-gap edge states. The yellow and green shaded sections represent the band regions of $l_{x1}=2.5$mm and $l_{x1}=4.5$mm, respectively. } \label{fig:10}
\end{figure}

The physical reason for the topological phase transition at $l_{x1}=3.6$mm can also be explained from two aspects. First, shown in the second and third columns of Fig. \ref{fig:9}(c), the gap width of the $1$st and $3$rd gap achieve minimum at $l_{x1}=3.6$mm. Second, we find that the intra-(inter-)cell spacing between upper rods is exactly the same as the inter-(intra-)cell spacing between lower rods, which coincides with the topological phase transition point $\delta=\Delta$, i.e. the inter-(intra-)cell hopping of the upper chain equals the intra-(inter-)cell hopping of the lower chain, of Rice-Mele ladder in the fractional-filling gaps discussed in Section \uppercase\expandafter{\romannumeral4}.

To confirm the different topological properties, we can consider the structure consisting of two types of PhCs with opposite topological phases to investigate the topologically protected edge states. As shown in Fig. \ref{fig:10}(a), eleven-cell PhC with $l_{x1}=2.5$mm is attached to eleven-cell PhC with $l_{x1}=4.5$mm. Three topological edge states appear in three gaps whose $E_z$ distributions are shown in Fig. \ref{fig:10}(b). The edge state in the $1$st gap shows clear evidence of coupling between $\begin{matrix}[1 &  1] \end{matrix}^T$ state which is similar to the edge state of Rice-Mele ladder in the $\frac{3}{4}$ filling gap shown in Fig. \ref{fig:6}(d). The edge state in the $3$rd gap exhibits the coupling between $\begin{matrix}[1 &  -1] \end{matrix}^T$ which is similar to the edge state of Rice-Mele ladder in the $\frac{1}{4}$ filling gap shown in Fig. \ref{fig:6}(c). On the other hand, the edge state in the $2$nd gap exhibits the zig-zag feature, which coincides with the edge state of Rice-Mele ladder in the half-filling gap shown in Fig. \ref{fig:2}(a). Thus, we conclude that both kinds of the reconstructed topology in Rice-Mele ladder can also realized in its photonic counterpart.

\subsection{\label{sec:level2}Experimental Observation of Topological Edge States in Photonic Rice-Mele Ladder}

\begin{figure}[t]
	\centering
	\includegraphics[width=1\linewidth]{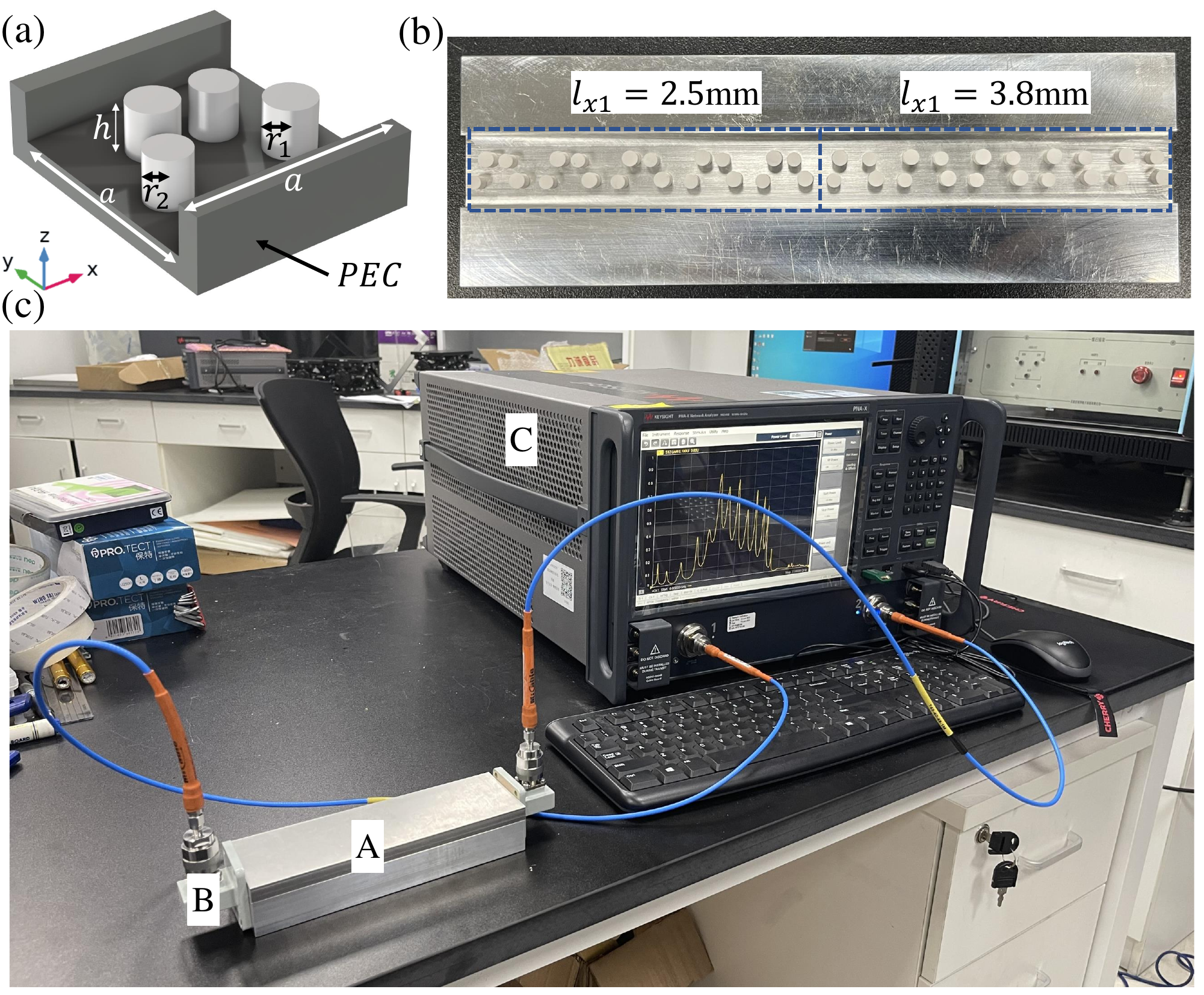}
	\caption{(a) Three-dimensional structure of a unit cell with the upper plate removed. (b) Photograph of the bonded PhC structure with the upper plate removed. The boundary of two topologically opposite structures with $N=5$ are marked by dashed boxes. (c) Photograph of the experimental setup for the transmission spectra. The combined PhC structure (A) is connected to the network analyzer (C) through waveguide-coaxial coupler (B). } \label{fig:11}
\end{figure}

To verify the effectiveness of our theory of topology reconstruction, here we try to observe the topological edge states through experiments at microwave frequency range from $6$ GHz to $11$ GHz. The three-dimensional (3D) realization of the photonic Rice-Mele ladder is shown in Fig. \ref{fig:11}(a). Here we set the height of dielectric rods $h=4$mm. Two boundaries in $z$ direction are taken as PEC so that the system is quite similar to the 2D model introduced in the last section. For frequencies under consideration, only $E_z$-polarized modes are available whose band-gap structure is the same as the 2D photonic Rice-Mele ladder introduced above.

To investigate the topological edge states, as shown in Fig. \ref{fig:11}(b), we consider the structure composed of two topologically opposite PhCs with $l_{x1}=2.5$mm and $l_{x1}=3.8$mm, respectively. Since the bonded PhCs can be regarded as a rectangle waveguide propagating in $x$ direction, the topological edge state can be observed in the form of transmission peak inside certain gap from transmission spectra. The experimental setup to detect the transmission spectra is shown in Fig. \ref{fig:11}(c). The ports of PNA-X Network Analyzer N5245B of Keysight are connected to the bonded PhC structure with waveguide-coaxial couplers to send and detect the signals, respectively. Since the waveguide-coaxial coupler can only activate the fundamental TE$10$ modes, here we would like to take the $1$st gap edge state to represent the perturbation-based topology and the $2$nd gap edge state to represent the isomorphic-mapping-based topology. To optimize the experimental results, for the frequencies around the $1$st gap, we set the number of unit cells $N=11$ with waveguide-coaxial coupler BJ70. For the frequencies around the $2$nd gap, we set the number of unit cells $N=6$ with waveguide-coaxial coupler BJ100.

\begin{figure}[t]
	\centering
	\includegraphics[width=1\linewidth]{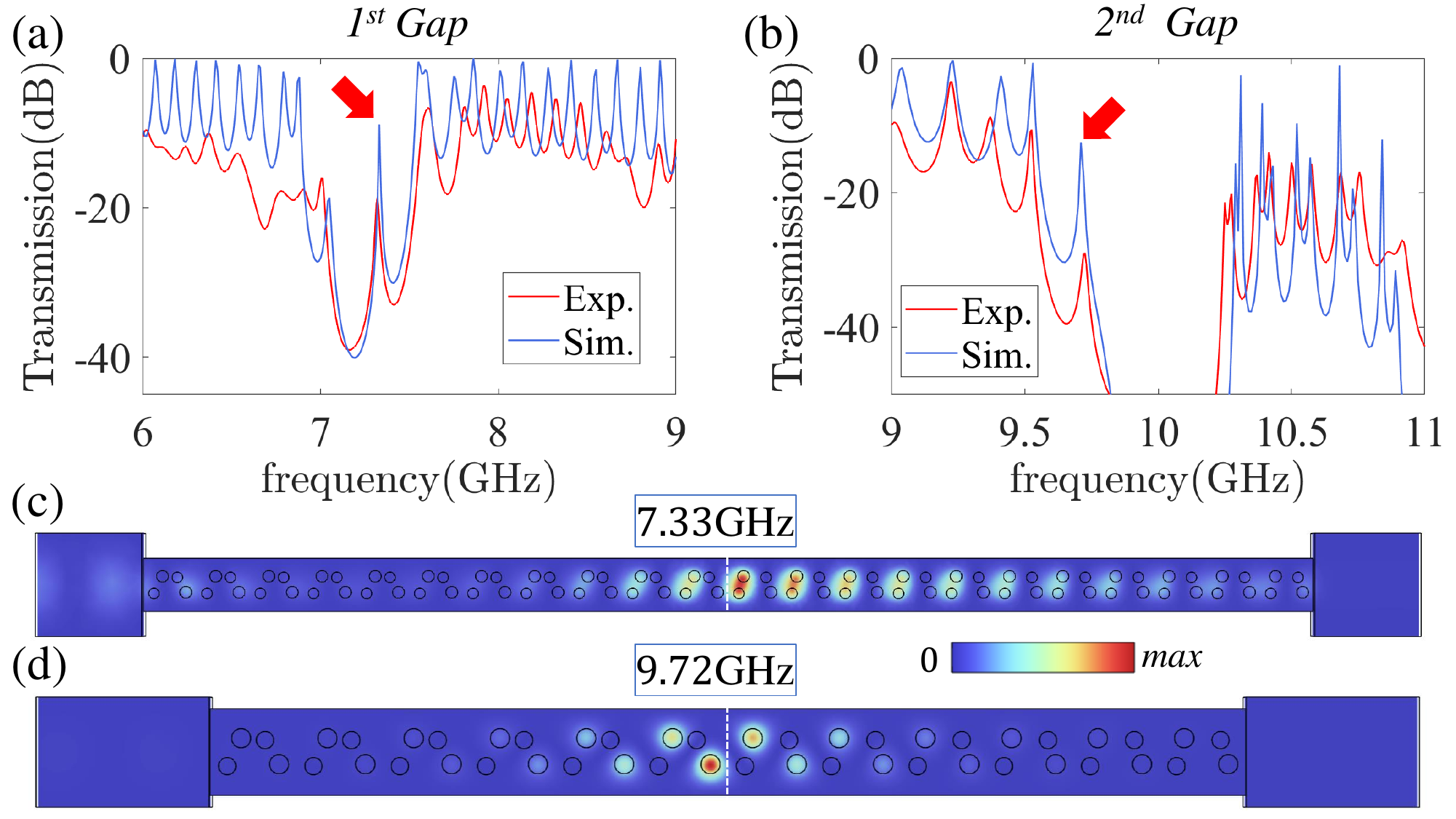}
	\caption{ (a-b) The transmission spectra of (a) the $1$st gap and (b) the $2$nd gap. The numerical and experimental results are denoted by the red lines and the blue lines, respectively. The in-gap transmission peaks are marked by red arrows. (c-d) The simulated $|E_z|^2$ distribution at the transmission peaks (c) $7.33$ GHz and (d) $9.72$ GHz, respectively. Both of them are highly localized at the boundary.} \label{fig:12}
\end{figure}

For the $1$st gap, the experimental transmission spectra are marked by red line in Fig. \ref{fig:12}(a), which agree very well with the numerical results marked by blue line. An in-gap transmission peak from the topological edge state can be found at $7.33$ GHz for both numerical and experimental results. The simulated field distribution $|E_z|^2$ at $7.33$ GHz is shown in Fig. \ref{fig:12}(c) exhibiting clear evidence of edge state defined on $\begin{matrix}[1 &  1] \end{matrix}^T$, same as theoretically predicted.

For the $2$nd gap, as shown in Fig. \ref{fig:12}(b), an in-gap transmission peak from the topological edge state can also be found near $9.72$ GHz for experiment (red line) and simulation (blue line). The zig-zag characteristic of the simulated $|E_z|^2$ distribution at $9.72$ GHz is shown in Fig. \ref{fig:12}(d) which coincides with our theoretical prediction.

One may notice that the absolute value of transmission spectra obtained by the experiment is smaller than the theoretical expectations despite sharing a similar trend. Several possible reasons may explain the difference, such as the absorption of the ceramic rods or the reflection at the boundary between our PhC structure and the waveguide-coaxial coupler. Nevertheless, we experimentally demonstrate that both of the topology reconstruction methods by the isomorphic mapping and the perturbation approximation can be realized in photonic systems.
	
\section{\label{sec:level1}CONCLUSION}
In summary, taking Rice-Mele-like systems as examples, we propose two methods of topology reconstruction for the multi-band systems without symmetries. For the first method, we find a strict one-to-one mapping relationship between asymmetric Rice-Mele-like system and its chiral counterpart, so that the topology protected by chiral symmetry can be reconstructed. For the second method, through perturbation approximation, the whole Hilbert space can be reduced into degenerate subspaces where Rice-Mele-like topology can be discovered. Based on these methods, the complicated topology origin for the edge states inside the near-zero-energy gap at half-filling and other gaps at fractional-filling can be revealed. It's also shown that both of the methods can be utilized in various asymmetric systems beyond Rice-Mele ladder model to investigate their topological properties, i.e. 2D Rice-Mele model and 1D superconducting systems. As to the effective topological invariant, we introduce the renormalized real-space winding number which can identify these new topological phase transitions by the integral jumping. Both methods are further extended to photonic asymmetric systems. By simplified photonic TBM, we can calculate the photonic renormalized real-space winding number and the topological phases are clearly shown. Finally, the topological edge states of the photonic Rice-Mele ladder at different gaps are observed through experiments, which match precisely with the numerical results from our two methods. Our findings offer a new perspective on topology in complex systems. First, the isomorphic mapping goes far beyond the conventional adiabatic deformation method, i.e. it can connect the topology of two systems with \emph{intrinsic} differences. For example, the topological critical point without gap closing of asymmetric system can be mapped into the gap-closing one of symmetric system. Second, the perturbation approximation, which is thought to be irrelevant to the topological properties generally, could take an important role in the topology study since it can generate simpler subspaces whose non-trivial topology could be revealed. From this view, we could have missed many topological phenomena if we had not gone deeper into the approximation subspace to explore their topological origin. Even more, these topological edge states from the approximation method show extraordinary robustness and can go far beyond the proper range of perturbation approximation, which implies that \emph{the strong perturbation can extend the original topology to the systems which might look very different}. In other words, approximation might lead to new classifications of topological insulators.

Our work leaves several open questions for further study. First, for the isomorphic mapping method of topology reconstruction, we need to choose \emph{proper} sublattices. Is there a general criterion of such choosing proper sublattices? Second, we show that the topological edge states from approximation-based topology still exist when the perturbation is relatively strong. So where is the boundary of perturbation-based topology? Furthermore, what's the mechanism behind the boundary? If these questions can be answered, we believe that the topology of asymmetric systems will be understood more thoroughly.

\begin{acknowledgments}
		
This work is supported by National High Technology Research and Development Program of China (17-H863-04-ZT-001-035-01); National Key Research
and Development Program of China (2016YFA0301103, 2018YFA0306201); National Natural Science Foundation of China (12174073). We are grateful to Rong Li and Yan Guo for useful discussions.
		
\end{acknowledgments}
	
\appendix
	
\section{The Necessary Condition for Real-Space Winding Number} \label{A}
	
In this appendix, we give a detailed proof on Eq. \eqref{eq6}. Although the result has been briefly mentioned in Ref. \cite{PhysRevB.103.224208,PhysRevLett.128.127601}, here we would like to emphasis that it is actually a consequence of chiral symmetry and should be regarded as a necessary condition for the validity of real-space winding number.

The Hamitonian of chiral system takes the form:
\begin{eqnarray} \label{A1}
	\mathcal{H}=\begin{bmatrix}
		0&h \\
		h^{\dagger} &0
	\end{bmatrix}.
\end{eqnarray}
Resemble the method used in the topology reconstruction of Rice-Mele-like system in Section \uppercase\expandafter{\romannumeral3}, the eigen-equation of $\mathcal{H}$ can be decomposed into two decoupled equations:
\begin{eqnarray} \label{A2}
	\left\{\begin{matrix}
		hh^{\dagger}\psi_{An}=E_n^2\psi_{An} \\
		h^{\dagger}h\psi_{Bn}=E_n^2\psi_{Bn}
	\end{matrix}\right. .
\end{eqnarray}
As a result, $\psi_A$ and $\psi_B$ are both orthogonal in their own subspaces:
\begin{eqnarray} \label{A3}
	\langle \psi_{An} | \psi_{Am} \rangle = \langle \psi_{Bn} | \psi_{Bm} \rangle = 0 ,\ \text{when}\ n \neq m.
\end{eqnarray}

On the other hand, as chiral operator $\Gamma$ anti-commutes with $\mathcal{H}$, for any eigenstate $\psi_n=\begin{bmatrix}\psi_{An} & \psi_{Bn}\end{bmatrix}^{T}$, $\Gamma\psi_n=\begin{bmatrix}\psi_{An} & -\psi_{Bn}\end{bmatrix}^{T}$ is also an eigenstate with opposite energy, which ensures that $\psi_n$ and $\Gamma\psi_n$ are nondegenerate states. From the orthogonality condition, we have:
\begin{eqnarray} \label{A4}
	\langle \psi_n | \Gamma\psi_n \rangle = \langle \psi_{An} | \psi_{An} \rangle - \langle \psi_{Bn} | \psi_{Bn} \rangle = 0 ,
\end{eqnarray}
Consider the normalization condition:
\begin{eqnarray} \label{A5}
	\langle \psi_n | \psi_n \rangle = \langle \psi_{An} | \psi_{An} \rangle + \langle \psi_{Bn} | \psi_{Bn} \rangle = 1,
\end{eqnarray}
together we get:
\begin{eqnarray} \label{A6}
	\langle \psi_{An} | \psi_{Bn} \rangle = \langle \psi_{Bn} | \psi_{Bn} \rangle = 0.5.
\end{eqnarray}
In terms of $U_\sigma$ matrix, Eq. \eqref{A6} and Eq. \eqref{A3} can be interpreted as:
\begin{eqnarray} \label{A7}
	U_AU_A^{\dagger}=U_A^{\dagger}U_A=U_BU_B^{\dagger}=U_B^{\dagger}U_B=0.5.
\end{eqnarray}

\section{Chiral Bloch Eigenstates at The Topological Phase Transition Point of Rice-Mele-Like Systems} \label{B}
In this appendix, we would like to prove that a pair of chiral Bloch eigenstates can be found if and only if the Rice-Mele-like system is at the topological phase transition point.

For any Rice-Mele-like system and its chiral counterpart:
\begin{eqnarray}
	\mathcal{H}_{RM}(k,R)=\begin{bmatrix} \label{B1}
		vI_n & h(k,R) \\
		h(k,R)^{\dagger} & -vI_n
	\end{bmatrix},\\ \mathcal{H}_{Chiral}(k,R)=\begin{bmatrix} \label{B2}
		0 & h(k,R) \\
		h(k,R)^{\dagger} & 0
	\end{bmatrix},
\end{eqnarray}
where $k$ is the Bloch vector, $R$ is the tuning parameter. Based on the isomorphic mapping relationship, both of them share the same topological phase transition point $R_C$. For $\mathcal{H}_{Chiral}$ at $R_C$, the gap around zero energy closes, i.e. there should be a pair of doubly degenerate zero-energy Bloch states $\psi_1(k_0),\psi_2(k_0)$. Considering the chiral operator $\Gamma$, one can see that $\Gamma\psi_1(k_0),\Gamma\psi_2(k_0)$ are also the zero-energy eigenstates:
\begin{eqnarray} \label{B3}
	\mathcal{H}_{Chiral}\Gamma\psi_1(k_0)=-\Gamma\mathcal{H}_{Chiral}\psi_1(k_0)=0.
\end{eqnarray}
Through the combination of degenerate eigenstates $\psi_1(k_0)$, $\psi_2(k_0)$, $\Gamma\psi_1(k_0)$, $\Gamma\psi_2(k_0)$, the zero-energy eigenstates of $\mathcal{H}_{Chiral}(k_0,R_C)$ can be taken as chiral states with opposite sublattices occupied:
\begin{eqnarray} \label{B4}
	\left\{\begin{matrix}
		\psi_1(k_0)'=\begin{bmatrix}
			\varphi_{An} \\
			0
		\end{bmatrix}
		\\\\
		\psi_2(k_0)'=\begin{bmatrix}
			0 \\
			\varphi_{Bn}
		\end{bmatrix}
	\end{matrix}\right. .
\end{eqnarray}
According to Eq. \eqref{eq16}, $\psi_1(k_0)',\psi_2(k_0)'$ are also the eigenstates of $\mathcal{H}_{RM}(k_0,R_C)$. Therefore, at the topological phase transition point of Rice-Mele-like system, there exist a pair of chiral Bloch eigenstates.

On the other hand, if $\mathcal{H}_{RM}(k,R)$ possesses a pair of chiral Bloch eigenstates, they must also be zero-energy eigenstates of its chiral counterpart $\mathcal{H}_{Chiral}(k,R)$. Therefore, for $\mathcal{H}_{Chiral}(k,R)$, the gap around zero energy closes, indicating both $\mathcal{H}_{Chiral}(k,R)$ and $\mathcal{H}_{RM}(k,R)$ are at the topological phase transition point $R=R_C$.

\section{Example of Direct Perturbation-Based Topology Reconstruction in 1D Ladder Systems} \label{C}

\begin{figure}[t]
	\centering
	\includegraphics[width=1\linewidth]{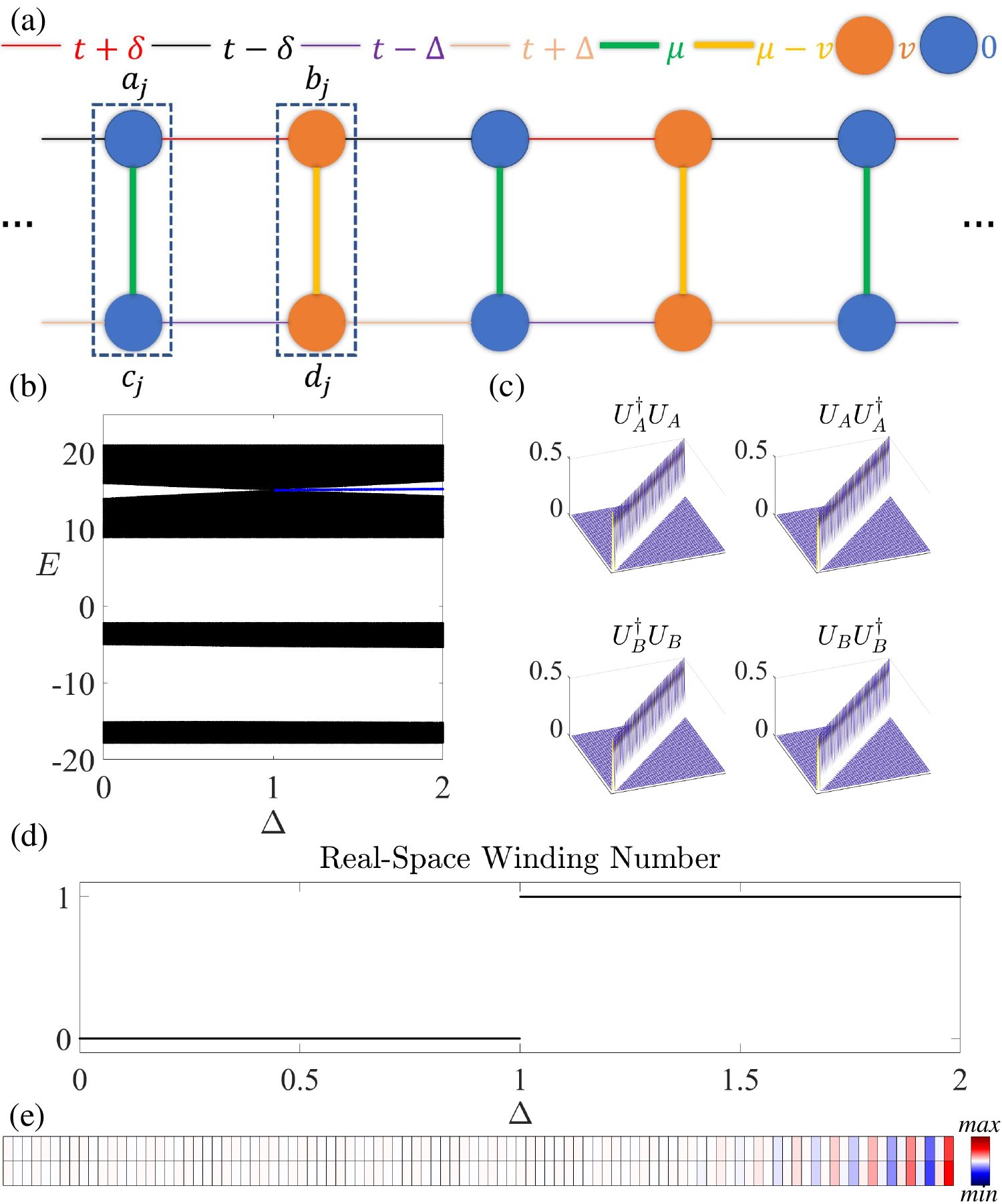}
	\caption{ (a) Schematic of the example 1D ladder model. The dimers in $y$ direction are marked by dashed boxes. (b) The band structure versus $\Delta$ with $t=3$, $\delta=1$, $\mu=15$, $v=5$. (c) The distributions of the four product matrices $U_AU_A^{\dagger}$, $U_A^{\dagger}U_A$, $U_BU_B^{\dagger}$, $U_B^{\dagger}U_B$ calculated by Eq. \eqref{eq36}. (d) The real-space winding number exhibits a unit jump at the critical point $\delta=\Delta$. (e) The spatial distribution of the edge state in the $3$rd gap. } \label{fig:13}
\end{figure}

To verify the independence of our perturbation-based topology, in this appendix we would like to exhibit an example of the 1D ladder model whose topology in the fractional-filling gaps can be obtained directly through degenerate perturbation.

The schematic of the 1D ladder model is shown in Fig. \ref{fig:13}(a), which could be realized by introducing additional onsite energies to the site $b$,$d$ of SSH ladder. Similar to Rice-Mele ladder, both spatial symmetry and chiral symmetry are broken here. The Hamiltonian can be written as:
\begin{eqnarray} \label{C1}
	\begin{aligned}
		\mathcal{H}&=\sum_{j}[(t+\delta)b_j^{\dagger}a_j+(t-\delta)a_{j+1}^{\dagger}b_j+(t-\Delta)d_j^{\dagger}c_j
		\\
		&+(t+\Delta)c_{j+1}^{\dagger}d_j+\mu a_j^{\dagger}c_j+(\mu-v) b_j^{\dagger}d_j+H.c.]
		\\
		&+vb_j^{\dagger}b_j+vd_j^{\dagger}d_j.
	\end{aligned}
\end{eqnarray}
Similarly, in the strong coupling limit $\mu$, $v \gg t$, $\delta$, $\Delta$, we consider the dimers composed of two atoms in $y$ direction:
\begin{eqnarray} \label{C2}
	\mathcal{H}_{dimer}^{A}=
	\begin{bmatrix}
		0 & \mu\\
		\mu & 0
	\end{bmatrix}, \mathcal{H}_{dimer}^{B}=
	\begin{bmatrix}
		v & \mu-v\\
		\mu-v & v
	\end{bmatrix} .
\end{eqnarray}
The eigen-energies and eigenstates of dimers can be expressed as:
\begin{eqnarray} \label{C3}
	\left\{\begin{matrix}
		E_{1i}^A = -\mu, \ \psi_{1i}^A = \frac{1}{\sqrt{2}}(|i,a\rangle - |i,c\rangle)
		\\
		E_{2i}^A = \mu, \ \psi_{2i}^A = \frac{1}{\sqrt{2}}(|i,a\rangle + |i,c\rangle)
		\\
		E_{1i}^B = -\mu+2v, \ \psi_{1i}^B = \frac{1}{\sqrt{2}}(|i,b\rangle - |i,d\rangle)
		\\
		E_{2i}^B = \mu, \ \psi_{2i}^B = \frac{1}{\sqrt{2}}(|i,b\rangle + |i,d\rangle)
	\end{matrix}\right. ,
\end{eqnarray}
where $\{\psi_{2i}^A\}$ and $\{\psi_{2i}^B\}$ share the same eigen-energy. Therefore, they form the degenerate subspace. The perturbation Hamiltonian of the degenerate subspace can be written as:
\begin{eqnarray} \label{C4}
	\begin{aligned}
		\mathcal{H}'&=\sum_{j}[t_1 | \psi_{2j}^A \rangle \langle \psi_{2j}^B | + t_2 | \psi_{2j}^B \rangle \langle \psi_{2(j+1)}^A |+H.c.],
	\end{aligned}
\end{eqnarray}
where $t_1=t+\frac{1}{2}(\delta-\Delta)$, $t_2=t-\frac{1}{2}(\delta-\Delta)$. One can easily see that $\mathcal{H}'$ is exactly the Hamiltonian of SSH model. Based on the conclusion of Section \uppercase\expandafter{\romannumeral4}, a pair of in-gap edge states should be found in the $3$rd gap around $E=\mu$ when $\delta<\Delta$, which is confirmed in the band structure shown in Fig. \ref{fig:13}(b). On the other hand, since the subspace perturbation Hamiltonian $\mathcal{H}'$ already possesses chiral symmetry, shown in Fig. \ref{fig:13}(c), the unitary matrices $U_A,U_B$ can be obtained directly via Eq. \eqref{eq36} without renormalization. The real-space winding number calculated by $U_A,U_B$ shown in Fig. \ref{fig:13}(d) is quantized and a unit jump can be observed at the critical point $\delta=\Delta$. Furthermore, the spatial distribution of the $3$rd gap edge state shown in Fig. \ref{fig:13}(e) exhibits clear evidence of chiral edge state defined on the degenerate subspace $\{\psi_{2i}^A\}$, $\{\psi_{2i}^B\}$. Therefore, in this example of 1D ladder model, its topological properties at $\frac{3}{4}$ filling can be reconstructed independently via perturbation theory.

\section{Perturbation-Based Topology in Dimerized Bi-Particle Superconducting System} \label{D}
In previous works on 1D bi-particle superconducting systems \cite{PhysRevB.98.024205,PhysRevResearch.3.013122,PhysRevB.90.014505}, it is found that the Hamiltonian of Kitaev chain in Majorana quasi-particle representation shares similar form with that of SSH ladder. Theoretically speaking, all the topological properties of SSH ladder should have counterparts in superconducting systems. As the Kitaev chain belongs to BD\uppercase\expandafter{\romannumeral1} class, only the topological properties at zero energy have been well discussed. Following the perturbtion-based topology reconstruction method introduced in the main text, in this appendix we would like to exhibit the new non-zero topological phase in 1D dimerized Kitaev chain, namely the Rice-Mele-like fermionic phase. This finding shows the generality of our method as well as its potential to greatly extend the study of topological band theory in various systems.

First, we transform the Bloch Hamiltonian of our non-superconducting model of Eq. \eqref{eq2} with $v=0$ into Majorana quasi-particle representation \cite{PhysRevResearch.3.013122,PhysRevB.90.014505}:
\begin{eqnarray} \label{D1}
	\mathcal{H}_{Maj}(k)=
	\begin{bmatrix}
		0 & [iM(k)]\\
		[iM(k)]^{\dagger} & 0
	\end{bmatrix}.
\end{eqnarray}
One can prove that $\mathcal{H}_{Maj}(k)$ and $\mathcal{H}(k)$ in Eq. \eqref{eq2} share the same band structure and band topology. Then we can express the Hamiltonian in the  Bogoliubov–de Gennes (BdG) form \cite{PhysRevB.90.014505} through the following unitary transformation:
\begin{eqnarray}
	\mathcal{H}_{BdG}(k)=U'\mathcal{H}_{Maj}(k)U'^{\dagger},\\  \label{D2} U'=\frac{1}{\sqrt{2}} \label{D3}
	\begin{bmatrix}
		1 && 0 && i & 0 \\
		0 && 1 && 0 & i \\
		1 && 0 && -i & 0 \\
		0 && 1 && 0 & -i
	\end{bmatrix}.
\end{eqnarray}
We note that $U'$ actually represents the relationship between majorana quasi-particle and fermion:
\begin{eqnarray} \label{D4}
\left\{\begin{matrix}
	\gamma_1=\frac{1}{\sqrt{2}}(c+c^{\dagger})
	\\
	\gamma_2=\frac{i}{\sqrt{2}}(c-c^{\dagger})
\end{matrix}\right. .
\end{eqnarray}
Then we have:
\begin{eqnarray} \label{D5}
	\mathcal{H}_{BdG}(k)=
	\begin{bmatrix}
		\mu_1 & z & 0 & w \\
		z^{*} & \mu_2 & -w^{*} & 0 \\
		0 & -w & -\mu_1 & -z \\
		w^{*} & 0 & -z^{*} & -\mu_2
	\end{bmatrix},
\end{eqnarray}
where $z=t_1+t_2e^{ik}$, $w=-\frac{1}{2}t'+\frac{1}{2}t'e^{ik}$, $t_1=t+\frac{1}{2}(\delta-\Delta)$, $t_2=t-\frac{1}{2}(\delta-\Delta)$, $t'=\delta+\Delta$. Finally, we obtain its real-space representation by performing inverse Fourier transformation to Eq. \eqref{D5}:
\begin{eqnarray} \label{D6}
	\begin{aligned}
		\mathcal{H}&=\sum_{n}(\mu_1c_{A,n}^{\dagger}c_{A,n}+\mu_2c_{B,n}^{\dagger}c_{B,n})
		\\
		&+(t_1c_{A,n}^{\dagger}c_{B,n}+t_2c_{A,n+1}^{\dagger}c_{B,n}+H.c.)
		\\
		&+\frac{1}{2}t'(c_{A,n}^{\dagger}c_{B,n}^{\dagger}+c_{A,n+1}^{\dagger}c_{B,n}^{\dagger}+H.c.).
	\end{aligned}
\end{eqnarray}
The Hamiltonian in Eq. \eqref{D6} is the superconducting counterpart of SSH ladder, which can be regarded as a dimerized Kitaev with staggered onsite energy. For the gap around zero energy, it is well-known that there exist a pair of Kitaev-like majorana modes with non-trivial winding number \cite{PhysRevB.98.024205}.

On the other hand, since Eq. \eqref{D6} and SSH ladder share the same band-gap structure, for the non-zero gaps, there should also be non-zero in-gap edge states when $t_2>t_1$. Their topological properties might be explained through our perturbation method. One may notice that the dimer representation presented in Section \uppercase\expandafter{\romannumeral4} is equivalent to the following unitary transformation:
\begin{eqnarray} \label{D7}
	U=\frac{1}{\sqrt{2}}
	\begin{bmatrix}
		1 && 0 && 1 & 0 \\
		0 && 1 && 0 & 1 \\
		1 && 0 && -1 & 0 \\
		0 && 1 && 0 & -1
	\end{bmatrix},
\end{eqnarray}
which shares a similar form with $U'$ in Eq. \eqref{D3}. Comparing Eq. \eqref{D3} with Eq. \eqref{D7}, we can obtain the following relationship:
\begin{eqnarray} \label{D8}
	\left\{\begin{matrix}
		c_{1i,A} \longrightarrow\ c_{A,i}^{\dagger}
		\\
		c_{2i,A} \longrightarrow\ c_{A,i}
		\\
		c_{1i,B} \longrightarrow\ c_{B,i}^{\dagger}
		\\
		c_{2i,B} \longrightarrow\ c_{B,i}
	\end{matrix}\right. ,
\end{eqnarray}
where $c_{1i,\sigma},c_{2i,\sigma}$ are the annihilation operators correspond to Eq. \eqref{eq24}:
\begin{eqnarray} \label{E9}
	\left\{\begin{matrix}
		c_{1i,A} = \frac{1}{\sqrt{2}}(c_{i,a} - c_{i,c})
		\\
		c_{2i,A} = \frac{1}{\sqrt{2}}(c_{i,a} + c_{i,c})
		\\
		c_{1i,B} = \frac{1}{\sqrt{2}}(c_{i,b} - c_{i,d})
		\\
		c_{2i,B} = \frac{1}{\sqrt{2}}(c_{i,b} + c_{i,d})
	\end{matrix}\right. .
\end{eqnarray}
That is to say, the degenerate subspace $\{\psi_{1i}^A, \psi_{1i}^B\}$ defined in Section \uppercase\expandafter{\romannumeral4} corresponds to the hole subspace in Eq. \eqref{D6} while subspace $\{\psi_{2i}^A, \psi_{2i}^B\}$ corresponds to particle subspace. From this point of view, the subspace Hamiltonian is equivalent to Eq. \eqref{D6} without the superconducting pairing gap $t'$, which is exactly the Hamiltonian of Rice-Mele model. Based on the conclusion of Section \uppercase\expandafter{\romannumeral3}, a pair of fermionic edge states with $E=\mu_1,\mu_2$ would appear when $t_2>t_1$, which coincides with the properties of non-zero edge states of SSH ladder at $\frac{3}{4}$ filling. Moreover, if we keep in mind the anti-communication relation between the annihilation operator $c$ and creation operator $c^\dagger$, in the hole subspace, the energies of fermionic edge states become $E=-\mu_1,-\mu_2$, which agree with the edge states of SSH ladder at $\frac{1}{4}$ filling. Therefore, from the perspective of Eq. \eqref{D6}, the perturbation-based topology of the $1$st gap and the $3$rd gap actually correspond to the same Rice-Mele-like fermionic topological phase in dimerized Kitaev chain. As a result, four rather than two non-zero edge states would appear in the topologically nontrivial region, which confirms its fermionic property as every fermion is divided into two quasi-particles in BdG form. It is worth noting that the Rice-Mele-like phase differs from the SSH-like phase protected by chiral symmetry in Ref. \cite{PhysRevB.90.014505} in two senses. First, it appears with non-zero $\mu_1,\mu_2$ and non-zero topological edge states; Second, it is independent of Kitaev-like topology. While the Kitaev-like topology protected by chiral symmetry (in the form of particle-hole symmetry for BD\uppercase\expandafter{\romannumeral1} superconducting systems) is determined by $t$ and $\mu_1,\mu_2$, the new Rice-Mele-like topology relies on the other two parameters $\delta$ and $\Delta$.

\nocite{*}
	
\bibliographystyle{apsrev4-2}
\bibliography{reference}% Produces the bibliography via BibTeX.
	
\end{document}